# Techno-Economic Planning of Spatially-Resolved Battery Storage Systems in Renewable-Dominant Grids Under Weather Variability

Seyed Ehsan **Ahmadi**[a], Elnaz **Kabir**[b], Mohammad **Fattahi**[a,*], Mousa **Marzband**[c,d] and Dongjun **Li**[a]

[a]*Newcastle Business School, Northumbria University, Newcastle upon Tyne, UK*
[b]*Department of Engineering Technology & Industrial Distribution, Texas A&M University, College Station, TX, USA*
[c]*Department of Electrical & Computer Engineering, King Abdulaziz University, Jeddah, Saudi Arabia*
[d]*Center of Research Excellence in Renewable Energy and Power Systems, Department of Electrical and Computer Engineering, King Abdulaziz University, Jeddah, Saudi Arabia*

ABSTRACT

The ongoing energy transition is significantly increasing the share of renewable energy sources (RES) in power systems; however, their intermittency and variability pose substantial challenges, including load shedding and system congestion. This study examines the role of the battery storage system (BSS) in mitigating these challenges by balancing power supply and demand. We optimize the location, size, and type of batteries using a two-stage stochastic program, with the second stage involving hourly operational decisions over an entire year. Unlike previous research, we incorporate the comprehensive technical and economic characteristics of battery technologies. The New York State (NYS) power system, currently undergoing a significant shift towards increased RES generation, serves as our case study. Using available load and weather data from 1980–2019, we account for the uncertainty of both load and RES generation through a sample average approximation approach. Our findings indicate that BSS can reduce renewable curtailment by 34% and load shedding by 21%, contributing to a more resilient power system in achieving NYS 2030 energy targets. Furthermore, the cost of employing BSS for the reduction of load shedding and RES curtailment does not increase linearly with additional capacity, revealing a complex relationship between costs and renewable penetration. This study provides valuable insights for the strategic BSS deployment to achieve a cost-effective and reliable power system in the energy transition as well as the feasibility of the NYS 2030 energy targets.

## 1. Introduction

The Paris Agreement emphasizes the urgency of limiting the global temperature rise to well below 2°C above pre-industrial levels to mitigate the risks and impacts of climate change [1]. Achieving this target requires an ambitious commitment to reaching net-zero greenhouse gas (GHG) emissions by the second half of the 21st century [2]. Since the power sector is a major source of GHG emissions, decarbonization through the integration of renewable energy sources (RES) is crucial [3]. In response, countries worldwide are advancing clean energy transition policies. For example, the European Union aims for at least 40% renewable energy penetration by 2030 [4], while China targets 35% by the same year [5]. The United States, as part of its broader plan to achieve net-zero emissions by 2050, aims to reduce GHG emissions by 50% below 2005 levels by 2030 and achieve full clean electricity by 2035 [6].

✉ seyed.ahmadi@northumbria.ac.uk (S.E. Ahmadi); ekabir@tamu.edu (E. Kabir); mohammad.fattahi@northumbria.ac.uk (M. Fattahi); mousa.marzband@gmail.com (M. Marzband); dongjun.li@northumbria.ac.uk (D. Li)
ORCID(s):

Decarbonizing the power system through large-scale renewable energy integration presents considerable challenges, primarily due to the intermittent nature of renewable generation. These challenges are further compounded by grid operational constraints, which lead to substantial energy curtailment, supply shortages, and transmission congestion [7, 8]. The battery storage system (BSS) emerges as a critical solution to these challenges. Without them, the installed capacity of variable renewable energy (VRE) can be overestimated, potentially undermining the efficiency of the energy transition [9]. As a multifaceted tool, batteries provide additional energy capacity [10], reduce energy curtailment, mitigate supply shortages, relieve transmission line congestion [11], stabilize the variability of RES [12], and manage peak load pressures [13].

Determining the optimal BSS capacity required to support a specific level of RES penetration (defined as the proportion of electricity generated from renewables in a power system) remains an open question that requires a systematic assessment [10] (Research Objective **1**). Effective BSS planning demands a thorough investigation, particularly given the diverse types of batteries available [14]. Each type of battery possesses unique technological and economic features that necessitate detailed evaluation [14, 15]. Furthermore, it is crucial to capture the interaction between these features and the grid's constraints, especially congestion [16]. This analysis must also employ a fine temporal scale [17], as coarser approaches, such as daily or weekly modelling, fail to reflect the real-time dynamics of batteries. Given the complexity of these variables and the constraints imposed by the power system at a fine temporal resolution, identifying an optimal solution constitutes a significant optimization challenge that remains insufficiently explored in existing literature (Research Objective **2**).

BSS planning must account for the variability of renewable generation and electricity load driven by weather fluctuations [18]. While scenario generation is often employed to address this, many studies lack spatial differentiation or fail to consider the co-variability of VRE generation and load. Effectively capturing the spatio-temporal variability and co-variability of key system parameters remains a critical challenge that has not been thoroughly addressed in existing research (Research Objective **3**). Furthermore, the impacts of BSS integration must be carefully evaluated to guide decision-makers on the costs and benefits of BSS. Existing metrics often focus on aggregated benefits, such as total system cost savings [17] or reliability improvements [14], without explicitly linking the costs of BSS integration to the benefits of addressing specific grid challenges. The literature underscores the need for actionable metrics that explicitly link the costs of BSS to their effectiveness in addressing these challenges (Research Objective **4**).

## 1.1. Practical Motivation

To demonstrate the practical effectiveness of the proposed battery storage optimization model, we apply it to the New York State (NYS) power system, a compelling real-world case study. Figure 1 illustrates the load zones in the New York State Control Area. In 2019, NYS enacted the Climate Leadership and Community Protection Act (CLCPA), which established some of the most ambitious climate goals in the country. These include achieving 70% renewable energy penetration and deploying 3,000 MW of bulk energy storage by 2030 [19]. Recognizing the critical role of energy storage in integrating VRE, reducing curtailment, and ensuring grid reliability, the Public Service Commission recently increased the energy storage target to 6,000 MW by 2030. This expansion is projected to save $2 billion in system costs while delivering significant public health benefits [20].

NYS also faces spatial imbalances, with renewable generation concentrated upstate and high electricity demand downstate, leading to transmission congestion, renewable curtailment, and load shedding. This diverse yet unevenly distributed power system provides a valuable context for evaluating the integration of various BSS technologies to comprehensively address these challenges. NYS's ambitious goals and complex grid dynamics make it an ideal case study for testing the model.

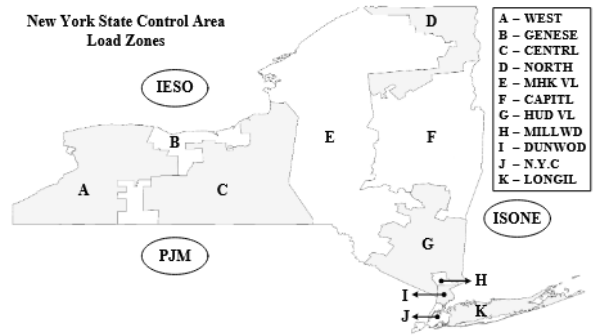

**Figure 1:** Formation of the 11 internal and 3 external load zones in the New York State Control Area.

## 1.2. Main Contributions

To address the research objectives, this paper proposes a joint optimization framework that integrates battery and grid constraints within a DC Optimal Power Flow (DC-OPF) formulation to minimize total system costs. DC-OPF is employed for two primary reasons: (i) its reduced data requirements compared to AC-OPF (particularly regarding voltage profiles and reactive power capabilities) and (ii) its computational efficiency, which is essential for solving large-scale, long-horizon problems.

The model incorporates hourly power dispatch across 8760 hours per year and evaluates a wide range of stochastic investment scenarios. It accounts for uncertainty in key parameters, including utility-scale wind power (UWT), utility-scale photovoltaic solar power (UPV), distributed solar power (DPV), big hydropower (BHD), small hydropower (SHD), and load. Given the dimensionality of these uncertainties and the extended operational horizon, solving an AC-based formulation would be computationally prohibitive. Instead, the DC-OPF formulation is embedded in a two-stage stochastic optimization model, solved using the Sample Average Approximation (SAA) method. This approach enables a tractable yet effective means of capturing the variability and uncertainty associated with RES and demand over long-term planning horizons.

We selected a stochastic framework rather than robust optimization (RO) for three key reasons.

1. The SAA method allows us to incorporate realistic uncertainty profiles derived from a 40-year historical dataset, preserving the co-variability among uncertain inputs such as wind speed, temperature, and solar radiation. This joint variability plays a critical role in energy balancing and storage performance but is difficult to model within RO, which typically relies on worst-case bounds and cannot readily capture statistical dependencies.

2. The scale of the problem (8760 hourly time steps and high-dimensional input uncertainty) makes RO computationally impractical. Finally, while RO is designed for risk-averse decision-making, it tends to produce overly conservative solutions in system-level planning, especially when historical data is available to inform more balanced, probabilistic scenario modelling.

3. The SAA-based stochastic optimization provides a more accurate and computationally viable approach for long-term strategic planning, enabling optimized investment decisions that are both data-driven and operationally realistic.

The NYS power system is also employed as a real-world case study, modified to align with the state's 2030 decarbonization targets under the CLCPA. These modifications include expanding transmission limits, integrating large-scale RES, and retiring fossil fuel and nuclear generators (FFGs and NPGs). The results provide actionable insights into BSS technology selection under diverse RES generation and load scenarios, offering valuable guidance for future energy transition planning. Accordingly, multiple managerial insights are presented and discussed.

The main contributions of this paper are summarized as follows:

- **Optimization for BSS sizing, technology selection, and grid integration:** In response to Research Objectives **1** and **2**, we propose a two-stage DC-OPF model. The first stage incorporates the capital, replacement, and fixed operation and maintenance (O&M) costs of BSS integration, while the second stage accounts for variable O&M costs, degradation costs of BSS, FFG generation, carbon emissions, and load shedding costs. By considering the technical and economic characteristics of six BSS technologies, the model optimizes the size and technology selection of BSS across the power system while ensuring compliance with all grid, dispatch, and transmission constraints.

- **Incorporating RES and load uncertainty via two-stage stochastic optimization and SAA:** In response to Research Objective **3**, we propose a two-stage stochastic programming model, solved using the SAA method. Hourly renewable generation and load samples are generated while explicitly capturing their spatio-temporal co-variability, utilizing historical weather data (e.g., wind speed, solar radiation, and temperature) and load patterns.

- **Novel metrics for measuring BSS cost-effectiveness:** In response to Research Objective **4**, we introduce two metrics to evaluate the cost-effectiveness of battery integration in reducing load shedding and RES curtailment. Load Shedding Reduction Cost-Effectiveness (LSRCE) quantifies the load-shedding reduction (MWh) per unit cost of BSS technologies, while Renewable Curtailment Reduction Cost-Effectiveness (RCRCE) measures RES curtailment reduction per unit cost of BSS technologies. These metrics are analysed across varying RES penetration rates and maximum allocated power capacities for the 2030 NYS power system.

The remainder of this paper is organized as follows. Section 2 reviews the literature on various solutions proposed for addressing large-scale VRE integration challenges, with a focus on BSS solutions. Section 3 defines the stochastic optimization problem for BSS allocation, accounting for both BSS and grid dispatch constraints. Section 4 details the 2030 NYS power system case study and its relevant data. Section 5 presents computational experiments and sensitivity analyses to derive managerial insights regarding BSS integration. Finally, Section 6 concludes the paper.

## 2. Literature Review

Transitioning to power systems with high renewable energy penetration requires addressing the inherent imbalance between electricity supply and demand caused by the variability and limited predictability of VRE. This short-term imbalance, occurring over minutes to days, poses a critical challenge as VRE adoption grows. To address this, various strategies have been explored and various solutions have been proposed, including non-VRE generation [21], fossil fuels with carbon capture [22], nuclear power, pumped storage, and demand-side management [10]. Among these, BSS has emerged as a key solution due to its versatile role in enhancing grid performance [17, 23]. BSS stabilizes renewable variability and mitigates supply shortages [12], reduces energy curtailment, relieves transmission congestion [11, 17], and manages peak loads [13, 24]. Additionally, they improve energy efficiency [25], lower electricity prices [26], and provide lifetime financial benefits [27], making them integral to achieving a reliable and efficient energy transition. Effective BSS planning requires careful consideration of multiple factors. The following reviews the literature through these lenses.

BSS exhibits diverse economic and technological characteristics that significantly influence their performance, efficiency, and suitability for different applications [15, 28, 29]. These characteristics can be broadly categorized into (i) cost, including capital and O&M expenses [14, 15, 28]; (ii) capacity and allocation, encompassing energy and power capacity, storage duration, and depth of charge/discharge [14, 18]; (iii) efficiency, such as round-trip efficiency and self-discharge rates [14, 15]; and (iv) lifetime and durability, which cover cycle life and degradation rates [14, 15, 28]. Other factors, such as environmental and sustainability impacts (e.g., life-cycle GHG emissions) [30], safety (e.g., thermal and fire risks), and reliability (e.g., forced outage rate) [28], are also noted but are often not mathematically modelled in planning studies. Interactions among these characteristics often result in trade-offs, such as cost versus efficiency, depth of discharge versus maximum cycle life, and self-discharge versus storage duration, complicating optimization. For instance, BSS technologies with lower efficiency, higher energy losses, shallow depth of discharge, or short discharge durations necessitate additional renewable generation and storage capacity, which in turn drives up total capital costs [15]. However, existing studies tend to simplify or overlook these trade-offs by focusing on average-case performance, limiting the realism and generalizability of their findings for long-term system planning.

Grid-level BSS technologies are categorized based on their technical, economic, and operational characteristics. Lithium-ion Batteries (LiB) are the most widely studied due to their technical maturity, as noted in [27] and [31]. In addition, Huang et al. [15] analysed five other technologies: Sodium-Sulfur Batteries (NaSB), Lead-Acid Batteries (LAB), Vanadium Redox Flow Batteries (VRFB), Nickel-Cadmium Batteries (NiCdB), and Zinc-Bromine Batteries (ZnBrB). Their study assumes a grid operating with one type of battery at a time, with each technology selected for its unique strengths. For instance, LiB and NiCdB are valued for maturity, NaSB and ZnBrB for scalability, LAB for low capital cost, and VRFB for superior cycling performance and long discharge durations. For the case study, they rank VRFB as the most preferred battery, while NiCdB is the least favoured due to inefficiencies and energy losses. Although most studies, such as [14, 17, 18, 28, 30], do not specify individual BSS technologies and instead focus on general storage considerations, results from [15] underscore the need to consider all relevant parameters in the optimal planning of BSS technologies, as overlooking these trade-offs can lead to suboptimal decisions regarding the size and type of BSS technologies for grid integration. There is a lack of comprehensive studies that simultaneously optimize across multiple BSS technologies while considering site-specific grid and renewable characteristics, which are crucial for practical deployment.

Battery characteristics, such as size and type, must align with grid constraints, particularly congestion and load patterns, to ensure efficient and reliable operations [16, 28]. Studies like [15, 17, 28] incorporate grid constraints, including transmission line limits and network operational conditions, into their analysis. Authors in [15] and [17] adopted a joint optimization approach that combines battery and grid constraints into a single problem to minimize total generation costs while simultaneously optimizing grid dispatch and battery operations. This method is effective for battery sizing, siting, and energy storage coordination with RES dispatch, congestion, and load management. In contrast, Peng et al. [28] employed a decoupled two-stage approach, using locational marginal prices from a grid dispatch problem as economic signals to guide battery operations. This approach is more suitable for maximizing economic benefits, such as battery arbitrage opportunities. While some studies explicitly account for grid constraints, many focus on system-wide battery planning without considering the spatial and operational complexities of the grid (e.g., [14, 18, 26, 27]). Ignoring these constraints can result in suboptimal deployment, increased costs, underutilized storage, and missed opportunities to enhance grid reliability. Moreover, there is a lack of studies that comprehensively integrate siting, sizing, technology selection, and grid modelling into a unified framework using long-term data.

In predominantly renewable grids, electricity supply fluctuates significantly across time, space, and scale, requiring meticulous planning [32, 33]. Several studies highlight the critical importance of accounting for this variability in BSS integration [15, 18]. For instance, Golombek et al. [31] used a two-stage stochastic programming approach to optimize investments in renewables, batteries, and transmission networks. However, they overlook spatio-temporal co-variability between renewable generation and load, which leads to suboptimal storage investments and sizing. This omission also misrepresents battery roles in renewable integration and distorts cost and operational analyses by focusing on average rather than extreme conditions. Alternatively, Jafari et al. [18] used the GenX model incorporating 11 years of historical renewable and load data to capture inter-annual variability in Italy's power system. However, the Italian power system is modelled as a single entity. While their approach is effective for capacity expansion and operational planning, its lack of spatial differentiation and transmission constraints limits accuracy in representing regional variability and the critical role of batteries in mitigating congestion. These gaps underscore the need for methodologies that fully capture spatio-temporal variability to optimize renewable integration and system reliability. There is a lack of studies that use large-scale historical data (e.g., across multiple decades) to guide battery planning models while capturing geographic diversity and uncertainty metrics.

In summary, while the literature provides valuable insights into BSS technologies, economic characteristics, grid integration, and uncertainty modelling, existing approaches remain fragmented. Most fail to simultaneously address multi-technology optimization, long-term weather and load variability, grid topology, and decision-value metrics within a unified stochastic planning framework.

## 3. Problem Formulation and Methodology

In this section, we formulate our problem as a two-stage stochastic program and propose the SAA algorithm to solve the optimization problem. The section is divided into five subsections: (3.1) characteristics of BSS technologies, (3.2) mathematical modelling of BSS, (3.3) a joint optimization model integrating BSS and grid constraints into a stochastic DC-OPF framework, (3.4) the SAA algorithm, and (3.5) the LSRCE and RCRCE metrics. Table 1 presents the nomenclature of the terms employed in the final optimization model. The additional terms specific to the initial BSS modelling are provided in Table 2.

### 3.1. Characteristics of BSS Technologies

In this section, the key performance characteristics of BSS technologies are defined as follows.

**Technical Characteristics**. *Rated Power Capacity* refers to the maximum instantaneous discharge power or the highest rate of discharge (in MW) that the BSS can achieve, starting from a fully charged state. *Energy Capacity* is the maximum amount of energy that can be stored (in MWh) [34]. *Storage Duration* is the time (in hours) that the BSS can discharge at its *Rated Power Capacity* before depleting its *Energy Capacity*. For example, a BSS with a *Rated Power Capacity* of 1 MW and a usable *Energy Capacity* of 4 MWh would have a *Storage Duration* of 4 hours. *Operational Lifetime/Maximum Number of Cycles* denotes the number of cycles or periods during which a BSS can undergo regular charging and discharging before experiencing failure or significant performance degradation. *Self-Discharge* refers to the loss of stored energy in a BSS due to internal chemical reactions, rather than active discharge to supply power to the grid, and is expressed as a percentage of lost charge over a specific period. *Depth of Discharge (DOD)* represents the percentage of the total energy capacity of a BSS that has been discharged relative to its full capacity. *Charging Efficiency* is the ratio of energy stored to energy supplied during the charging process, while *Discharging Efficiency* is the ratio of energy delivered by the BSS to the energy stored. *Round-Trip Efficiency (RTE)* measures the ratio of energy stored in a BSS to the energy retrieved from it. Lastly, the *Annual RTE Degradation Factor* quantifies the yearly reduction in the RTE of a BSS due to usage.

**Economic Characteristics.** *Capital Cost* of a BSS comprises four key components: 1) *Energy Capacity Cost*, which includes costs for electrodes, electrolytes, and separators; 2) *Power Conversion System Cost*, covering the costs of the inverter, packaging, and container; 3) *Balance of Plant Cost*, which encompasses costs for components such as site wiring, interconnecting transformers, and other ancillary equipment; and 4) *Construction and Commissioning Cost*, including costs for site design, equipment procurement/transportation, and labour/parts for installation. *Replacement Cost* refers to the total expenditure required to replace a BSS at the end of its useful lifetime or when it becomes non-functional. Also, *Fixed and Variable O&M Costs* account for all expenses needed to maintain the operational status of a BSS throughout its economic lifetime, irrespective of energy usage. Lastly, *Degradation Cost* is the financial implications of wear and performance degradation in a BSS over time due to usage.

## 3.2. Mathematical Modelling of BSS

**Energy Balance and Charging/Discharging**. Constraints (1)–(5) define the energy balance and charging/discharging dynamics of a BSS type $n \in N$ at bus $i \in I_z$ in zone $z \in Z$ for scenario $s \in S$. These constraints are adapted from [35], which defines linear formulations for energy balance and charging/discharging operations without relying on binary variables. Constraints (1) and (2) ensure the hourly energy balance and the adherence to the maximum/minimum allowable energy levels in each BSS, respectively. Constraint (3) indicates that the energy level of the BSS at the end of the time horizon ($T$) must equal its initial energy level (at time 0). Constraints (4) and (5) specify the maximum allowable charging and discharging values for the BSS at each time interval.

$$e^{bs}_{z,n,i,t,s} = e^{bs}_{z,n,i,t-1,s} + \left(\eta^{ch}_n \cdot p^{ch}_{z,n,i,t,s} \cdot \Delta t\right) - \left(\frac{p^{dc}_{z,n,i,t,s} \cdot \Delta t}{\eta^{dc}_n}\right) - \left(\frac{\eta^{sd}_n}{24} \cdot e^{bs}_{z,n,i,t-1,s}\right); \quad (1)$$

$$\forall z \in Z, n \in N, i \in I_z, t \in T, s \in S$$

$$\underline{e}^{bs}_{z,n,i} \leq e^{bs}_{z,n,i,t,s} \leq \overline{e}^{bs}_{z,n,i}; \quad \forall z \in Z, n \in N, i \in I_z, t \in T, s \in S \quad (2)$$

$$e^{bs}_{z,n,i,0,s} = e^{bs}_{z,n,i,T,s}; \quad \forall z \in Z, n \in N, i \in I_z, s \in S \quad (3)$$

$$0 \leq p^{ch}_{z,n,i,t,s} \leq \overline{p}^{ch}_{z,n,i}; \quad \forall z \in Z, n \in N, i \in I_z, t \in T, s \in S \quad (4)$$

$$0 \leq p^{dc}_{z,n,i,t,s} \leq \overline{p}^{dc}_{z,n,i}; \quad \forall z \in Z, n \in N, i \in I_z, t \in T, s \in S \quad (5)$$

where $p^{ch}_{z,n,i,t,s}$ and $p^{dc}_{z,n,i,t,s}$ represent the charging and discharging power of BSS type $n \in N$ at bus $i \in I_z$ in zone $z \in Z$ at time interval $\Delta t$ for scenario $s$, respectively. $\overline{p}^{ch}_{z,n,i}$ and $\overline{p}^{dc}_{z,n,i}$ denote the maximum charging and discharging values of BSS type $n \in N$ at bus $i \in I_z$ in zone $z \in Z$, respectively. Similarly, $\underline{e}^{bs}_{z,n,i}$ and $\overline{e}^{bs}_{z,n,i}$ represent the minimum and maximum energy stored in the BSS type $n \in N$ at bus $i \in I_z$ in zone $z \in Z$, respectively.

Unlike the formulation in [35], which seeks an optimal solution based on complementarity between charging and discharging, our approach adopts a non-complementary strategy presented by [36] to prevent simultaneous charging and discharging without relying on binary variables or complementarity constraints. In addition to our non-complementary strategy, the variable O&M and degradation costs naturally discourage simultaneous charging and discharging by penalizing it in the objective function. These costs are defined based on hourly charging and discharging values to improve the practical representation of BSS operation without binary constraints.

By dividing each time period $t$ into a charging subinterval $\Delta t^{ch}$ and a discharging subinterval $\Delta t^{dc}$, complementarity can be preserved as long as these subintervals do not overlap. Under this assumption, the model can be reformulated as follows:

Table 1: Nomenclature (Final Model)

**Sets and Indices**

| | |
|---|---|
| $Z$ | Set of zones indexed by $z \in Z = \{A, B, C, D, E, F, G, H, I, J, K, IESO, ISONE, PJM\}$, the zones are shown in Figure 1. |
| $I$ | Set of buses indexed by $i \in I$; $I_z$ is the set of buses in zone $z$ |
| $R$ | Set of RES indexed by $r \in R = \{BHD, SHD, UPV, DPV, UWT\}$; $R_z$ is the set of RES in zone $z$ |
| $G$ | Set of thermal power generators indexed by $g \in G = \{FFG, NPG\}$; $G_z$ is the set of thermal power generators in zone $z$ |
| $L$ | Set of transmission lines indexed by $l \in L = \{L_z^{in}, L_z^{ex}\}$; $L_z^{in}$ is the set of transmission lines connecting two buses in zone $z$ and $L_z^{ex}$ is the set of transmission lines connecting buses in zone $z$ to buses in other zones. |
| $N$ | Set of BSS types indexed by $n \in N = \{NaSB, LiB, LAB, ZEBRA, ZnBrB, VRFB\}$ |
| $T$ | Set of operating time indexed by $t \in T$ |
| $S$ | Set of scenarios indexed by $s \in S$ |

**Parameters**

| | |
|---|---|
| $\delta_n^{ec}$ | Energy capacity cost in capital cost of BSS type $n$ [\$/MWh] |
| $\delta_n^{pc}$ | Power conversion system cost in capital cost of BSS type $n$ [\$/MW] |
| $\delta_n^{bp}$ | Balance of plant cost in capital cost of BSS type $n$ [\$/MW] |
| $\delta_n^{cc}$ | Construction and commissioning cost in capital cost of BSS type $n$ [\$/MWh] |
| $\delta_n^{br}$ | Replacement cost of BSS type $n$ [\$/MW] |
| $\delta_n^{bf}$ | Fixed O&M cost of BSS type $n$ [\$/MW] |
| $\delta_n^{bv}$ | Variable O&M cost of BSS type $n$ [\$/MWh] |
| $\delta_n^{bd}$ | Degradation cost of BSS type $n$ [\$/MWh] |
| $\varphi^{bs}$ | Planned BSS capacity allocation [MW] |
| $\eta_n^{ch}, \eta_n^{dc}$ | Charging/Discharging efficiency of BSS type $n$ [%] |
| $\eta_n^{df}$ | Annual round-trip efficiency degradation factor of BSS type $n$ [%] |
| $\eta_n^{rt}$ | Round-trip efficiency of BSS type $n$ [%] |
| $\eta_n^{sd}$ | Daily self-discharge rate of BSS type $n$ [%] |
| $\overline{D}_n$ | Maximum depth-of-discharge of BSS type $n$ [%] |
| $\tilde{D}$ | Determined depth-of-discharge of BSS to calculate equivalent number of cycles [%] |
| $\upsilon_n$ | Storage duration (maximum discharge time at rated power) of BSS type $n$ [hours] |
| $K_n^{cl}$ | Maximum number of cycles for BSS type $n$ to calculate equivalent number of cycles |
| $K_n^{lt}$ | Operational lifetime of BSS type $n$ to calculate equivalent number of cycles and annualized replacement cost [year] |
| $K^{pt}$ | Time horizon for BSS installation planning to calculate annualized capital cost [year] |
| $\overline{P}_{z,r,i,t,s}^{re}$ | Maximum RES generation in zone $z$ of resource $r$ at bus $i$ at time $t$ for scenario $s$ [MW] |
| $\underline{P}_{z,g,i}^{dg}, \overline{P}_{z,g,i}^{dg}$ | Minimum/Maximum thermal power generation in zone $z$ of source $g$ at bus $i$ [MW] |
| $P_{z,g,i}^{ru}/P_{z,g,i}^{rd}$ | Ramp up/down of thermal generators in zone $z$ of source $g$ at bus $i$ [MW] |
| $\delta_{z,g,i,t}^{dg}$ | Cost of thermal generator in zone $z$ of source $g$ at bus $i$ at time $t$ [\$/MWh] |
| $\delta_{z,g,i}^{em}$ | $CO_2$ emission rate of fossil-based units in zone $z$ of source $g$ at bus $i$ [ton $CO_2$/MWh] |
| $\delta^{co}$ | Cost of carbon emission of fossil-based units [\$/ton $CO_2$] |
| $P_{z,i,t,s}^{ld}$ | Load in zone $z$ at bus $i$ at time $t$ for scenario $s$ [MW] |
| $\underline{P}_{z,l}^{fl}, \overline{P}_{z,l}^{fl}$ | Minimum/Maximum allowed power flow in zone $z$ for transmission line $l$ [MW] |
| $\underline{P}_z^{ex}, \overline{P}_z^{ex}$ | Minimum/Maximum allowed power exchange to zone $z$ [MW] |
| $B_{i,j}$ | Susceptance of transmission line connecting bus $i$ to bus $j$ [$ohm^{-1}$] |
| $in$ | Interest rate for the investment |
| $O$ | A large numerical value |

## Variables

| | |
|---|---|
| $C^{cc}_{z,n,i}$ | Annualized capital cost of BSS type $n \in N$ at bus $i \in I_z$ in zone $z \in Z$ [\$] |
| $C^{rc}_{z,n,i}$ | Annualized replacement cost of BSS type $n \in N$ at bus $i \in I_z$ in zone $z \in Z$ [\$] |
| $C^{fo}_{z,n,i}$ | Annual fixed O&M cost of BSS type $n \in N$ at bus $i \in I_z$ in zone $z \in Z$ [\$] |
| $C^{vo}_{z,n,i,s}$ | Annual variable O&M cost of BSS type $n \in N$ at bus $i \in I_z$ in zone $z \in Z$ for scenario $s \in S$ [\$] |
| $C^{bd}_{z,n,i,s}$ | Annual degradation cost of BSS type $n \in N$ at bus $i \in I_z$ in zone $z \in Z$ for scenario $s \in S$ [\$] |
| $\overline{p}^{bs}_{z,n,i}$ | Maximum allocated rated power capacity of BSS type $n$ at bus $i$ in zone $z$ [MW] |
| $p^{ch}_{z,n,i,t,s}$ | Charge value of BSS type $n$ at bus $i$ in zone $z$ at time $t$ for scenario $s$ [MW] |
| $p^{dc}_{z,n,i,t,s}$ | Discharge value of BSS type $n$ at bus $i$ in zone $z$ at time $t$ for scenario $s$ [MW] |
| $e^{bs}_{z,n,i,t,s}$ | Amount of energy stored in BSS type $n$ at bus $i$ in zone $z$ at time $t$ for scenario $s$ [MWh] |
| $K^{eq}_{z,n,i,s}$ | Maximum annual equivalent number of cycles of BSS type $n \in N$ at bus $i \in I_z$ in zone $z \in Z$ for scenario $s$ |
| $p^{re}_{z,r,i,t,s}$ | Renewable generation in zone $z$ of resource $r$ at bus $i$ at time $t$ for scenario $s$ [MW] |
| $p^{dg}_{z,g,i,t,s}$ | Thermal power generation in zone $z$ of source $g$ at bus $i$ at time $t$ for scenario $s$ [MW] |
| $p^{ls}_{z,i,t,s}$ | Amount of load shedding in zone $z$ at bus $i$ at time $t$ for scenario $s$ [MW] |
| $p^{fl}_{z,l,t,s}$ | Power flow of transmission line $l$ in zone $z$ at time $t$ for scenario $s$ [MW] |
| $\theta_{i,t,s}$ | Voltage phase angle at bus $i$ at time $t$ for scenario $s$ [°] |
| $\lambda^{bs}_s$ | Annual operational-planning cost of BSS for scenario $s$ [\$] |
| $\phi^{rc}_s, \phi^{ls}_s$ | Annual RES curtailment/load shedding reduction for scenario $s$ [MWh/year] |
| $\pi^{rc}, \pi^{ls}$ | Cost-effectiveness of reducing RES curtailment/load shedding [MWh-year/\$] |

Table 2: Nomenclature (Initial Formulations, for Reference)

## Parameters

| | |
|---|---|
| $\Delta t$ | Time interval [hours] |
| $\Delta t^{ch}, \Delta t^{dc}$ | Charging/Discharging subinterval [hours] |

## Variables

| | |
|---|---|
| $\overline{p}^{ch}_{z,n,i}$ | Maximum charging value of BSS type $n \in N$ at bus $i \in I_z$ in zone $z \in Z$ [MW] |
| $\overline{p}^{dc}_{z,n,i}$ | Maximum discharging value of BSS type $n \in N$ at bus $i \in I_z$ in zone $z \in Z$ [MW] |
| $\tilde{p}^{ch}_{z,n,i,t,s}$ | Charge value of BSS type $n \in N$ at bus $i \in I_z$ in zone $z \in Z$ at subintervals $\Delta t^{ch}$ and $\Delta t^{dc}$ for scenario $s$ [MW] |
| $\tilde{p}^{dc}_{z,n,i,t,s}$ | Discharge value of BSS type $n \in N$ at bus $i \in I_z$ in zone $z \in Z$ at subintervals $\Delta t^{ch}$ and $\Delta t^{dc}$ for scenario $s$ [MW] |
| $\underline{e}^{bs}_{z,n,i}$ | Minimum energy stored in BSS type $n \in N$ at bus $i \in I_z$ in zone $z \in Z$ [MWh] |
| $\overline{e}^{bs}_{z,n,i}$ | Maximum energy stored in BSS type $n \in N$ at bus $i \in I_z$ in zone $z \in Z$ [MWh] |

$$e^{bs}_{z,n,i,t,s} = e^{bs}_{z,n,i,t-1,s} + \left(\eta^{ch}_n \cdot \tilde{p}^{ch}_{z,n,i,t,s} \cdot \Delta t^{ch}\right) - \left(\frac{\tilde{p}^{dc}_{z,n,i,t,s} \cdot \Delta t^{dc}}{\eta^{dc}_n}\right) - \left(\frac{\eta^{sd}_n}{24} \cdot e^{bs}_{z,n,i,t-1,s}\right); \quad (6)$$
$$\forall z \in Z, n \in N, i \in I_z, t \in T, s \in S$$

$$\underline{e}^{bs}_{z,n,i} \leq e^{bs}_{z,n,i,t,s} \leq \overline{e}^{bs}_{z,n,i}; \quad \forall z \in Z, n \in N, i \in I_z, t \in T, s \in S \quad (7)$$

$$e^{bs}_{z,n,i,0,s} = e^{bs}_{z,n,i,T,s}; \quad \forall z \in Z, n \in N, i \in I_z, s \in S \quad (8)$$

$$0 \leq \tilde{p}^{ch}_{z,n,i,t,s} \leq \overline{p}^{ch}_{z,n,i}; \quad \forall z \in Z, n \in N, i \in I_z, t \in T, s \in S \quad (9)$$

$$0 \leq \tilde{p}^{dc}_{z,n,i,t,s} \leq \overline{p}^{dc}_{z,n,i}; \quad \forall z \in Z, n \in N, i \in I_z, t \in T, s \in S \tag{10}$$

$$\Delta t^{ch}, \Delta t^{dc} \in \mathbb{R}^+, \quad \Delta t^{ch} + \Delta t^{dc} \leq \Delta t; \quad \forall t \in T \tag{11}$$

where $\tilde{p}^{ch}_{z,n,i,t,s}$ and $\tilde{p}^{dc}_{z,n,i,t,s}$ represent the charging and discharging power of BSS type $n \in N$ at bus $i \in I_z$ in zone $z \in Z$ at subintervals $\Delta t^{ch}$ and $\Delta t^{dc}$ for scenario $s$, respectively. Constraint (11) ensures that the charging and discharging subintervals in period $t$, $\Delta t^{ch}$ and $\Delta t^{dc}$, do not overlap. However, this model is nonlinear due to the bilinear terms $\tilde{p}^{ch}_{z,n,i,t,s} \cdot \Delta t^{ch}$ and $\tilde{p}^{dc}_{z,n,i,t,s} \cdot \Delta t^{dc}$ in Constraint (6). Recall the relaxed formulation presented in Constraints (1)–(5), consider a given pair of $\tilde{p}^{ch}_{z,n,i,t,s}$ and $\tilde{p}^{dc}_{z,n,i,t,s}$, along with the corresponding energy level. If $\tilde{p}^{ch}_{z,n,i,t,s} \cdot \tilde{p}^{dc}_{z,n,i,t,s} > 0$, it is necessary to verify whether this configuration is executable. we can assess the feasibility of this pair by verifying the existence of $\Delta t^{ch}$ and $\Delta t^{dc}$ that satisfies Constraints (6) and (11). To this end, we must identify a pair of charging and discharging strategies, $p^{ch}_{z,n,i,t,s}$ and $p^{dc}_{z,n,i,t,s}$, along with the time intervals, $\Delta t^{ch}$ and $\Delta t^{dc}$, such that $\Delta t^{ch} + \Delta t^{dc} \leq \Delta t$ and the energy level in period $t \in T$ remains the same. To ensure this, the following condition, as outlined in Equations (12) and (13), must be satisfied.

$$\tilde{p}^{ch}_{z,n,i,t,s} \cdot \Delta t^{ch} = p^{ch}_{z,n,i,t,s} \cdot \Delta t; \quad \forall z \in Z, n \in N, i \in I_z, t \in T, s \in \tag{12}$$

$$\tilde{p}^{dc}_{z,n,i,t,s} \cdot \Delta t^{dc} = p^{dc}_{z,n,i,t,s} \cdot \Delta t; \quad \forall z \in Z, n \in N, i \in I_z, t \in T, s \in S \tag{13}$$

To limit charging and discharging time and avoid overlap between subintervals, Constraints (9) and (10) can be adjusted by incorporating the maximum allowed charging and discharging values, $\overline{p}^{ch}_{z,n,i}$ and $\overline{p}^{dc}_{z,n,i}$, as specified in Equation (13). This adjustment results in the definition of Constraints (14) and (15).

$$\tilde{p}^{ch}_{z,n,i,t,s} \leq \overline{p}^{ch}_{z,n,i} \rightarrow \frac{p^{ch}_{z,n,i,t,s} \cdot \Delta t}{\overline{p}^{ch}_{z,n,i}} \leq \Delta t^{ch}; \quad \forall z \in Z, n \in N, i \in I_z, t \in T, s \in S \tag{14}$$

$$\tilde{p}^{dc}_{z,n,i,t,s} \leq \overline{p}^{dc}_{z,n,i} \rightarrow \frac{p^{dc}_{z,n,i,t,s} \cdot \Delta t}{\overline{p}^{dc}_{z,n,i}} \leq \Delta t^{dc}; \quad \forall z \in Z, n \in N, i \in I_z, t \in T, s \in S \tag{15}$$

Constraint (16) is derived by summing Constraints (14) and (15) to ensure non-overlapping subintervals.

$$\frac{p^{ch}_{z,n,i,t,s} \cdot \Delta t}{\overline{p}^{ch}_{z,n,i}} + \frac{p^{dc}_{z,n,i,t,s} \cdot \Delta t}{\overline{p}^{dc}_{z,n,i}} \leq \Delta t^{ch} + \Delta t^{dc} \leq \Delta t; \quad \forall z \in Z, n \in N, i \in I_z, t \in T, s \in S \tag{16}$$

or equivalently, Constraint (17), which is independent of the duration of time $t \in T$.

$$\frac{p^{ch}_{z,n,i,t,s}}{\overline{p}^{ch}_{z,n,i}} + \frac{p^{dc}_{z,n,i,t,s}}{\overline{p}^{dc}_{z,n,i}} \leq 1; \quad \forall z \in Z, n \in N, i \in I_z, t \in T, s \in S \tag{17}$$

Based on the definitions of *Rated Power Capacity*, denoted as $\overline{p}^{bs}_{z,n,i}$, and *Energy Capacity*, denoted as $\overline{e}^{bs}_{z,n,i}$, provided in Section 3.1, we have $\overline{e}^{bs}_{z,n,i} = v_n \cdot \overline{p}^{bs}_{z,n,i}$ and $\underline{e}^{bs}_{z,n,i} = (1 - \overline{D}_n) \cdot v_n \cdot \overline{p}^{bs}_{z,n,i}$. Also, $\overline{p}^{ch}_{z,n,i}$ and $\overline{p}^{dc}_{z,n,i}$ are assumed to be equal to $\overline{p}^{bs}_{z,n,i}$. Thus, the linear mathematical model of BSS can be described by Constraints (18)–(22), considering that the time interval $\Delta t$ is set to 1 hour.

$$e^{bs}_{z,n,i,t,s} = e^{bs}_{z,n,i,t-1,s} + \left( \eta^{ch}_n \cdot p^{ch}_{z,n,i,t,s} - \frac{p^{dc}_{z,n,i,t,s}}{\eta^{dc}_n} \right) - \left( \frac{\eta^{sd}_n}{24} \cdot e^{bs}_{z,n,i,t-1,s} \right); \tag{18}$$
$$\forall z \in Z, n \in N, i \in I_z, t \in T, s \in S$$

$$(1 - \overline{D}_n) \cdot v_n \cdot \overline{p}^{bs}_{z,n,i} \leq e^{bs}_{z,n,i,t,s} \leq v_n \cdot \overline{p}^{bs}_{z,n,i}; \quad \forall z \in Z, n \in N, i \in I_z, t \in T, s \in S \tag{19}$$

$$e^{bs}_{z,n,i,0,s} = e^{bs}_{z,n,i,T,s}; \quad \forall z \in Z, n \in N, i \in I_z, s \in S \tag{20}$$

$$p^{ch}_{z,n,i,t,s} + p^{dc}_{z,n,i,t,s} \leq \overline{p}^{bs}_{z,n,i}; \quad \forall z \in Z, n \in N, i \in I_z, t \in T, s \in S \tag{21}$$

$$p^{ch}_{z,n,i,t,s}, p^{dc}_{z,n,i,t,s} \in \mathbb{R}^+ \tag{22}$$

**Equivalent Number of Cycles.** The formulation of the equivalent number of cycles is provided in Constraint (23) for BSS type $n \in N$ at bus $i \in I_z$ in zone $z \in Z$ for scenario $s \in S$.

$$K^{eq}_{z,n,i,s} = \frac{1}{2} \cdot \frac{\sum_{t \in T} \left| e^{bs}_{z,n,i,t,s} - e^{bs}_{z,n,i,t-1,s} \right|}{\tilde{D} \cdot \overline{e}^{bs}_{z,n,i}}; \quad \forall z \in Z, n \in N, i \in I_z, s \in S \tag{23}$$

Given that the maximum annual equivalent number of cycles is constrained by the operational lifetime ($K^{lt}_n$) and the maximum number of cycles ($K^{cl}_n$), Constraint (23) can be reformulated as in Constraint (24). In this paper, the value of $\tilde{D}$ is defined as 80%. Besides, as we provided a non-complementary strategy to prevent simultaneous charging and discharging of BSS, the absolute value energy level change is equal to the charging and discharging values. Thus, we have $|e^{bs}_{z,n,i,t,s} - e^{bs}_{z,n,i,t-1,s}| = p^{ch}_{z,n,i,t,s} + p^{dc}_{z,n,i,t,s}$.

$$K^{eq}_{z,n,i,s} \leq \frac{K^{cl}_n}{K^{lt}_n} \rightarrow \frac{1}{2} \cdot \frac{\sum_{t \in T} \left( p^{ch}_{z,n,i,t,s} + p^{dc}_{z,n,i,t,s} \right)}{\tilde{D} \cdot v_n \cdot \overline{p}^{bs}_{z,n,i}} \leq \frac{K^{cl}_n}{K^{lt}_n}; \quad \forall z \in Z, n \in N, i \in I_z, s \in S \tag{24}$$

It should be noted that the DOD of the BSS at the beginning and end of the year is set to be equal. This condition implies that $\sum_{t \in T} p^{ch}_{z,n,i,t,s} = \sum_{t \in T} p^{dc}_{z,n,i,t,s}$. Accordingly, we have:

$$\sum_{t \in T} p^{dc}_{z,n,i,t,s} \leq \frac{K^{cl}_n \cdot \tilde{D} \cdot v_n \cdot \overline{p}^{bs}_{z,n,i}}{K^{lt}_n}; \quad \forall z \in Z, n \in N, i \in I_z, s \in S \tag{25}$$

**Annualized Capital and Replacement Cost.** The annualized capital cost ($C^{cc}_{z,n,i}$) and replacement cost ($C^{rc}_{z,n,i}$) of a BSS type $n \in N$ at bus $i \in I_z$ in zone $z \in Z$ are calculated per unit of *Rated Power Capacity* (in \$/MW). Considering the *Storage Duration* of the BSS ($v_n$), the annualized capital cost is determined based on the time horizon for BSS installation planning, as expressed in Equation (26). Similarly, the annualized replacement cost is calculated based on the operational lifetime of the BSS, as outlined in Equation (27).

$$C^{cc}_{z,n,i} = \left( \delta^{ec}_n \cdot v_n + \delta^{pc}_n + \delta^{bp}_n + \delta^{cc}_n \cdot v_n \right) \cdot \left[ \frac{in \cdot (in + 1)^{K^{pt}}}{(in + 1)^{K^{pt}} - 1} \right] \cdot \overline{p}^{bs}_{z,n,i}; \quad \forall z \in Z, n \in N, i \in I_z \tag{26}$$

$$C^{rc}_{z,n,i} = \delta^{br}_n \cdot \upsilon_n \cdot \left[\frac{in}{(in+1)^{K^{lt}_n} - 1}\right] \cdot \overline{p}^{bs}_{z,n,i}; \quad \forall z \in Z, n \in N, i \in I_z \tag{27}$$

where the term $\frac{in \cdot (in+1)^{K^{pt}}}{(in+1)^{K^{pt}} - 1}$ represents the annualized capital cost factor, which evenly distributes the capital cost of the BSS across its operational lifetime. Similarly, the term $\frac{in}{(in+1)^{K^{lt}_n} - 1}$ represents the annualized replacement cost factor, which allocates the replacement cost of the BSS over its installation planning period.

**Annual Fixed and Variable O&M Cost.** The annual fixed O&M cost ($C^{fo}_{z,n,i}$) of a BSS type $n \in N$ at bus $i \in I_z$ in zone $z \in Z$ is defined and calculated as shown in Equation (28).

$$C^{fo}_{z,n,i} = \delta^{bf}_n \cdot \overline{p}^{bs}_{z,n,i}; \quad \forall z \in Z, n \in N, i \in I_z \tag{28}$$

Similarly, the annual variable O&M cost ($C^{vo}_{z,n,i,s}$) of a BSS type $n \in N$ at bus $i \in I_z$ in zone $z \in Z$ for scenario $s \in S$ is defined and calculated as shown in Equation (29).

$$C^{vo}_{z,n,i,s} = \sum_{t \in T} \delta^{bv}_n \cdot \left| e^{bs}_{z,n,i,t,s} - e^{bs}_{z,n,i,t-1,s} \right|; \quad \forall z \in Z, n \in N, i \in I_z, s \in S \tag{29}$$

The charging and discharging values of the BSS are equal to the absolute value of the energy level change as we have a non-simultaneous charging and discharging in BSS model. Accordingly, Equation (29) can be reformulated by incorporating the charging and discharging values within each time interval in Equation (30).

$$C^{vo}_{z,n,i,s} = \sum_{t \in T} \delta^{bv}_n \cdot \left( p^{ch}_{z,n,i,t,s} + p^{dc}_{z,n,i,t,s} \right); \quad \forall z \in Z, n \in N, i \in I_z, s \in S \tag{30}$$

**Annual Degradation Cost.** The annual degradation cost ($C^{bd}_{z,n,i,s}$) of a BSS type $n \in N$ at bus $i \in I_z$ in zone $z \in Z$ for scenario $s \in S$ represents the cost associated with the annual RTE degradation factor, as defined in Equation (31). This cost quantifies the decline in system performance due to battery aging and other factors, enabling an evaluation of long-term operational viability.

$$C^{bd}_{z,n,i,s} = \sum_{t \in T} \delta^{bd}_n \cdot \frac{\eta^{df}_n}{\eta^{rt}_n} \cdot \left| e^{bs}_{z,n,i,t,s} - e^{bs}_{z,n,i,t-1,s} \right|; \quad \forall z \in Z, n \in N, i \in I_z, s \in S \tag{31}$$

Here, the term $|e^{bs}_{z,n,i,t,s} - e^{bs}_{z,n,i,t-1,s}|$ represents the absolute change in energy level over consecutive time periods and can be used when full cycle data isn't available, as supported by Xu et al. [37]. The battery degradation cost per MWh, $\delta^{bd}_n$, is similar to the marginal cost of cycling proposed in [37]. The ratio $\eta^{df}_n / \eta^{rt}_n$ is the key innovation of this formulation, which accounts for round-trip efficiency degradation and supports the practical findings presented in [38] and [39]. The parameters used in this ratio is sourced from the PNNL Energy Storage Technology and Cost Characterization Report [40].

This approach serves as an approximation that captures battery aging effects over time without detailed electrochemical modelling, making it suitable for long-term system planning. The degradation factor reflects the impact of cycle life and depth of discharge, so more frequent or deeper cycling leads to higher degradation costs. This allows performance decline to be accounted for in investment and dispatch decisions, without modelling the physical degradation process directly.

As we have $|e^{bs}_{z,n,i,t,s} - e^{bs}_{z,n,i,t-1,s}| = p^{ch}_{z,n,i,t,s} + p^{dc}_{z,n,i,t,s}$, Equation (31) is reformulated as in Equation (32).

$$C^{bd}_{z,n,i,s} = \sum_{t \in T} \delta^{bd}_n \cdot \frac{\eta^{df}_n}{\eta^{rt}_n} \cdot \left(p^{ch}_{z,n,i,t,s} + p^{dc}_{z,n,i,t,s}\right); \quad \forall z \in Z, n \in N, i \in I_z, s \in S \tag{32}$$

### 3.3. Stochastic Optimization Model Integrating BSS and Grid Constraints

The BSS allocation problem is modelled as a two-stage stochastic program. The uncertain parameters in this formulation are the load and the maximum renewable energy generation from BHD, SHD, UPV, DPV, and UWT. These are scenario-based, generated from multiple years of historical weather data. The vector of these parameters is denoted by $\xi$. For this problem, $\xi = (\boldsymbol{P^{ld}}, \boldsymbol{\overline{P}^{re}})$, where $\xi^s$ represents a specific realization of the uncertain parameters in scenario $s \in S$. The objective function seeks to minimize the total annual operational-planning cost of the power system in the two-stage stochastic program, as expressed in Equation (33). It should be noted that **bold** parameters/variables indicate that some or all indices have been eliminated, resulting in a reduced-dimensional representation.

$$\min \sum_{z \in Z} \sum_{n \in N} \sum_{i \in I_z} \left[C^{cc}_{z,n,i} + C^{rc}_{z,n,i} + C^{fo}_{z,n,i}\right] + \mathbb{E}\left[Q\left(\boldsymbol{\overline{p}^{bs}}, \xi\right)\right] \tag{33}$$

Subject to:

$$\sum_{z \in Z} \sum_{n \in N} \sum_{i \in I_z} \overline{p}^{bs}_{z,n,i} \leq \varphi^{bs} \tag{34}$$

$$\boldsymbol{\overline{p}^{bs}} \in \mathbb{R}^+ \tag{35}$$

The model determines the optimal location, size, and technology of each BSS through the first-stage decision variable $\overline{p}^{bs}_{z,n,i}$, which represents the maximum rated power capacity of BSS type $n$ at bus $i$ in zone $z$. This variable simultaneously encodes (i) location via the zone and bus indices $z$ and $i$, (ii) technology type via the index $n$, which distinguishes six BSS technologies (see Table 4), and (iii) size via the continuous variable value $\overline{p}^{bs}_{z,n,i}$. These investment decisions are made before the realization of uncertainty and passed into the second-stage problem to evaluate hourly operational performance across multiple stochastic scenarios.

The function $Q(\boldsymbol{\overline{p}^{bs}}, \xi^s)$ is formulated as follows:

$$\min \; Q(\boldsymbol{\overline{p}^{bs}}, \xi^s) = \sum_{z \in Z} \sum_{n \in N} \sum_{i \in I_z} \left[C^{vo}_{z,n,i,s} + C^{bd}_{z,n,i,s}\right] + \sum_{z \in Z} \sum_{i \in G_z} \sum_{t \in T} \left[\left(\delta^{dg}_{z,g,i,t} + \delta^{co} \cdot \delta^{em}_{z,g,i}\right) \cdot p^{dg}_{z,g,i,t,s}\right] \\ + \sum_{z \in Z} \sum_{i \in I_z} \sum_{t \in T} \left[O \cdot p^{ls}_{z,i,t,s}\right] \tag{36}$$

Subject to:

Constraints (18)-(21) and (25)

$$p^{re}_{z,r,i,t,s} \leq \overline{P}^{re}_{z,r,i,t,s}; \quad \forall z \in Z, r \in R, i \in R_z, t \in T \tag{37}$$

$$\underline{P}^{dg}_{z,g,i} \leq p^{dg}_{z,g,i,t,s} \leq \overline{P}^{dg}_{z,g,i}; \quad \forall z \in Z, g \in G, i \in G_z, t \in T \tag{38}$$

$$p^{dg}_{z,g,i,t,s} - p^{dg}_{z,g,i,t-1,s} \leq P^{rd}_{z,g,i}; \quad \forall z \in Z, g \in G, i \in G_z, t \in T \tag{39}$$

$$p^{dg}_{z,g,i,t-1,s} - p^{dg}_{z,g,i,t,s} \leq P^{ru}_{z,g,i}; \quad \forall z \in Z, g \in G, i \in G_z, t \in T \tag{40}$$

$$p^{ls}_{z,i,t,s} \leq P^{ld}_{z,i,t,s}; \quad \forall z \in Z, i \in I_z, t \in T \tag{41}$$

$$\underline{P}^{fl}_{z,l} \leq p^{fl}_{z,l,t,s} \leq \overline{P}^{fl}_{z,l}; \quad \forall z \in Z, l \in L^{in}_z, t \in T \tag{42}$$

$$\underline{P}^{ex}_z \leq \sum_{l \in L^{ex}_z} p^{fl}_{z,l,t,s} \leq \overline{P}^{ex}_z; \quad \forall z \in Z, t \in T \tag{43}$$

$$p^{fl}_{z,l,t,s} = B_{i,j} \left( \theta_{i,t,s} - \theta_{j,t,s} \right); \quad \forall z \in Z, l \in L, i \in I_z, j \in I_z, t \in T \tag{44}$$

$$\sum_{i \in I_z} P^{ld}_{z,i,t,s} - \sum_{i \in I_z} p^{ls}_{z,i,t,s} - \sum_{l \in L^{ex}_z} p^{fl}_{z,l,t,s} = \\ \sum_{i \in G_z} p^{dg}_{z,g,i,t,s} + \sum_{i \in R_z} p^{re}_{z,r,i,t,s} + \sum_{n \in N} \sum_{i \in I_z} \left( p^{ch}_{z,n,i,t,s} - p^{dc}_{z,n,i,t,s} \right); \quad \forall z \in Z, t \in T \tag{45}$$

$$p^{ch}, p^{dc}, p^{re}, p^{ls} \in \mathbb{R}^+ \tag{46}$$

The cost components in Equation (33) include the annualized capital, replacement, and fixed O&M costs of the BSS, which are defined based on the maximum allocated capacity, as detailed in Subsection 3.2. Constraint (34) ensures that the total allocated capacity of the BSS be less than or equal to the system's planned capacity. Constraint (35) enforces the non-negativity of the allocated BSS capacity. Equation (36) represents the objective function of the second stage, which accounts for the variable O&M costs, degradation costs of the BSS, operating costs of thermal power generators, carbon emissions costs, and load-shedding costs. These costs are scenario-dependent and vary based on the uncertainty realization represented by the scenarios. Constraint (37) ensures that the generation of RES remains within their allowable capacity. Constraints (38)–(40) represent the generation and ramp-rate limits of thermal power generators. Constraint (41) defines the maximum allowable load shedding. Constraints (42)–(44) specify the power flow limits for each transmission line. Constraint (45) ensures hourly power balance in each zone. Finally, Constraint (46) enforces the non-negativity of the decision variables.

### 3.4. Sample Average Approximation Method

The two-stage stochastic programming model incorporates uncertain parameters related to hourly electricity load and maximum renewable generation. These parameters cannot be accurately characterized using straightforward stochastic processes due to their spatial-temporal correlations and variability throughout the year. To address this, the study uses 40 years of historical weather data (1980–2019) to predict maximum renewable generation, combined with corresponding load data, to construct 40 stochastic scenarios, as described in Section 4. However, the optimization model becomes computationally intractable when considering all 40 scenarios. To reduce the problem size, the SAA algorithm is employed. In the SAA method, $M'$ scenarios are randomly selected from the 40 scenarios $W$ times, and the stochastic program is solved optimally for each of these $W$ sets. The optimal objective values obtained from these $W$ solutions are used to estimate the mean ($\mu_L$) and variance ($\sigma^2_L$) of a lower bound (LB) for the true objective function. Subsequently, the $(1 - \alpha)\%$ confidence interval (CI) of the lower bound is computed as follows:

$$\left[ \mu_L - t_{\alpha/2, W-1} \frac{\sigma_L}{\sqrt{W}}, \mu_L + t_{\alpha/2, W-1} \frac{\sigma_L}{\sqrt{W}} \right] \tag{47}$$

Next, the $W$ feasible solutions (first-stage decisions) obtained are fixed in the optimization problem, which is then solved for $M$ scenarios. In other words, these solutions are simulated using $M$ scenarios, where $M > M'$. In this study, $M = 40$. The average of the simulation responses for each solution provides an upper bound (UB) for the true objective function. Finally, the solution with the minimum average of simulation responses is selected as the final decision and is used to estimate the upper bound of the problem's true objective. The main steps of the SAA algorithm are outlined in Algorithm 1. Additional details on the SAA method can be found in [41] and [42].

---

**Algorithm 1** The SAA method

---

**Step 1:** Selecting random Scenarios
**for** $w = 1, 2, \ldots, W$ **do**
    Select randomly $M'$ scenarios from the set of $M$ scenarios.
    Solve the two-stage stochastic program with selected $M'$ scenarios
    Achieve optimal objective $h^w$ and first-stage decisions $x^w$.
    **for** scenarios $s = 1, 2, \ldots, M$ **do**
        Fix the first-stage solution $x^w$ and solve the problem for scenario $s$.
        Achieve the optimal objective for solution $x^w$ and scenario $s$ as $f^{w,s}$.
    **end for**
    Calculate the simulation response $r^w$ as $\frac{\sum_{s=1}^{M} f^{w,s}}{M}$.
**end for**
**Step 2:** Calculate the mean and variance for the lower bound of the true objective value:
$$\mu_L = \frac{\sum_{w=1}^{W} h^w}{W}, \quad \sigma_L^2 = \frac{1}{W-1} \sum_{w=1}^{W} (h^w - \mu_L)^2$$
The $(1-\alpha)\%$ CI for the lower bound is:
$$\left[ \mu_L - t_{\alpha/2, W-1} \frac{\sigma_L}{\sqrt{W}}, \mu_L + t_{\alpha/2, W-1} \frac{\sigma_L}{\sqrt{W}} \right]$$
**Step 3:** Calculate the mean and variance for the upper bound of the true objective value:
$$\mu_U = \min_{w=1,2,\ldots,W} (r^w)$$
$\hat{w}$ is the index related to $\min_{w=1,2,\ldots,W} (r^w)$, and $\sigma_U^2 = \frac{1}{M-1} \sum_{s=1}^{M} (f^{\hat{w},s} - \mu_U)^2$
The $(1-\alpha)\%$ CI for the upper bound is:
$$\left[ \mu_U - z_{\alpha/2} \frac{\sigma_U}{\sqrt{M}}, \mu_U + z_{\alpha/2} \frac{\sigma_U}{\sqrt{M}} \right]$$
**Step 4:** Return the final solution $x^{\hat{w}}$ for the first-stage decisions.

---

### 3.5. Cost-Effectiveness of Load Shedding and Renewable Curtailment Reduction

In this paper, optimal BSS allocation is proposed as the primary strategy for reducing RES curtailment and load shedding in the power system. To this end, the impact of the BSS operational-planning cost on reducing RES curtailment and load shedding is analysed in detail by comparing the results of the optimal BSS allocation with those of a model without BSS. The results of the optimal BSS allocation are denoted with a hat symbol ($\bar{\hat{p}}_{z,n,i}^{bs}, \hat{p}_{z,n,i,t,s}^{ch}, \hat{p}_{z,n,i,t,s}^{dc}, \hat{p}_{z,i,t,s}^{ls}, \hat{p}_{z,g,i,t,s}^{dg}, \hat{p}_{z,r,i,t,s}^{re}$), while the results of the model without BSS ($\varphi^{bs} = 0$) are indicated by a check symbol ($\check{p}_{z,i,t,s}^{ls}, \check{p}_{z,g,i,t,s}^{dg}, \check{p}_{z,r,i,t,s}^{re}$).

The annual operational-planning cost of BSS ($\lambda_s^{bs}$), is defined as the difference between the annual cost of the model with optimal BSS allocation and that of the model without BSS, as expressed in Equation (48). The annual RES curtailment reduction in MWh/year ($\phi_s^{rc}$) is defined as the difference between the annual RES curtailment in the optimal BSS allocation model and that in the model without BSS. This can also be expressed as the difference between the annual RES generation in the optimal BSS allocation model ($\hat{p}_{z,r,i,t,s}^{re}$) and that in the model without BSS ($\check{p}_{z,r,i,t,s}^{re}$), as shown in Equation (49). Similarly, the annual load shedding reduction in MWh/year ($\phi_s^{ls}$) is formulated as the difference between the annual load shedding in the optimal BSS allocation model ($\hat{p}_{z,i,t,s}^{ls}$) and that in the model without BSS ($\check{p}_{z,i,t,s}^{ls}$), as expressed in Equation (50).

$$\begin{aligned}
\lambda_s^{bs} = &\sum_{z \in Z}\sum_{n \in N}\sum_{i \in I_z} \left[ (\delta_n^{ec} \cdot v_n + \delta_n^{pc} + \delta_n^{bp} + \delta_n^{cc} \cdot v_n) \cdot \left[ \frac{in \cdot (in+1)^{K^{pt}}}{(in+1)^{K^{pt}} - 1} \right] \cdot \overline{\hat{p}}_{z,n,i}^{bs} \right] \\
&+ \sum_{z \in Z}\sum_{n \in N}\sum_{i \in I_z} \left[ \delta_n^{br} \cdot v_n \cdot \left[ \frac{in}{(in+1)^{K_n^{lt}} - 1} \right] \cdot \overline{\hat{p}}_{z,n,i}^{bs} \right] + \sum_{z \in Z}\sum_{n \in N}\sum_{i \in I_z} \left[ \delta_n^{bf} \cdot \overline{\hat{p}}_{z,n,i}^{bs} \right] \\
&+ \sum_{z \in Z}\sum_{n \in N}\sum_{i \in I_z}\sum_{t \in T} \left[ \left(\delta_n^{bv} + \delta_n^{bd} \cdot \frac{\eta_n^{df}}{\eta_n^{rt}}\right) \cdot \left(\hat{p}_{z,n,i,t,s}^{ch} + \hat{p}_{z,n,i,t,s}^{dc}\right) \right] \\
&+ \sum_{z \in Z}\sum_{i \in G_z}\sum_{t \in T} \left[ \left(\delta_{z,g,i,t}^{dg} + \delta^{co} \cdot \delta_{z,g,i}^{em}\right) \cdot \hat{p}_{z,g,i,t,s}^{dg} \right] \\
&- \sum_{z \in Z}\sum_{i \in G_z}\sum_{t \in T} \left[ \left(\delta_{z,g,i,t}^{dg} + \delta^{co} \cdot \delta_{z,g,i}^{em}\right) \cdot \check{p}_{z,g,i,t,s}^{dg} \right]; \quad \forall s \in S
\end{aligned} \quad (48)$$

$$\phi_s^{rc} = \sum_{z \in Z}\sum_{i \in R_z}\sum_{t \in T} \left[ \hat{p}_{z,r,i,t,s}^{re} - \check{p}_{z,r,i,t,s}^{re} \right]; \quad \forall s \in S \quad (49)$$

$$\phi_s^{ls} = \sum_{z \in Z}\sum_{i \in I_z}\sum_{t \in T} \left[ \check{p}_{z,i,t,s}^{ls} - \hat{p}_{z,i,t,s}^{ls} \right]; \quad \forall s \in S \quad (50)$$

To better analyse the impact of BSS, two novel metrics are introduced to measure the cost-effectiveness of reducing RES curtailment and load shedding. The RCRCE, (in MWh-year/$), is defined as the average amount of RES curtailment reduction per unit cost of BSS allocation. Similarly, the LSRCE (in MWh-year/$), is defined as the average amount of load-shedding reduction per unit cost of BSS allocation.

$$\pi^{rc} = \frac{1}{|S|} \cdot \sum_{s \in S} \frac{\phi_s^{rc}}{\lambda_s^{bs}} \quad (51)$$

$$\pi^{ls} = \frac{1}{|S|} \cdot \sum_{s \in S} \frac{\phi_s^{ls}}{\lambda_s^{bs}} \quad (52)$$

Equations (51) and (52) quantify the economic benefits of BSS allocation. A higher ratio reflects a more cost-effective solution, indicating that the BSS reduces RES curtailment and load shedding more efficiently relative to its investment. For clarity and interpretability, the units and definitions of these cost-effectiveness metrics are summarized in Table 3.

## 4. 2030 NYS Power System Case Study

To achieve the goals of the CLCPA, the NYS Climate Action Council requires that 70% of statewide electricity come from renewable energy sources by 2030. Specific targets include installing 6000 MW of solar power by 2025, deploying 3000 MW of BSS by 2030, and adding 9000 MW of offshore UWTs by 2035 [6]. In support of the CLCPA, the New York Power Grid Study is undertaken to identify these upgrades for upstate (zones A to E) and downstate (zones F to K), as illustrated in Figure 2 [43]. It shows that upstate loads would be fully (100%) supplied by zero-emission resources, while downstate loads would achieve an 80% zero-emission supply by 2030.

Table 3: Summary of Cost-Effectiveness Metrics

| Metric | Units | Meaning |
|---|---|---|
| RCRCE ($\pi^{rc}$) | MWh-year/$ | How much renewable curtailment is avoided per $1 of BSS cost. Higher values indicate greater cost-effectiveness, reflecting more efficient utilization of RES and higher returns on investment. |
| LSRCE ($\pi^{ls}$) | MWh-year/$ | How much load shedding is avoided per $1 of BSS cost. Higher values indicate greater cost-effectiveness, reflecting more efficient supply of load and higher returns on investment. |

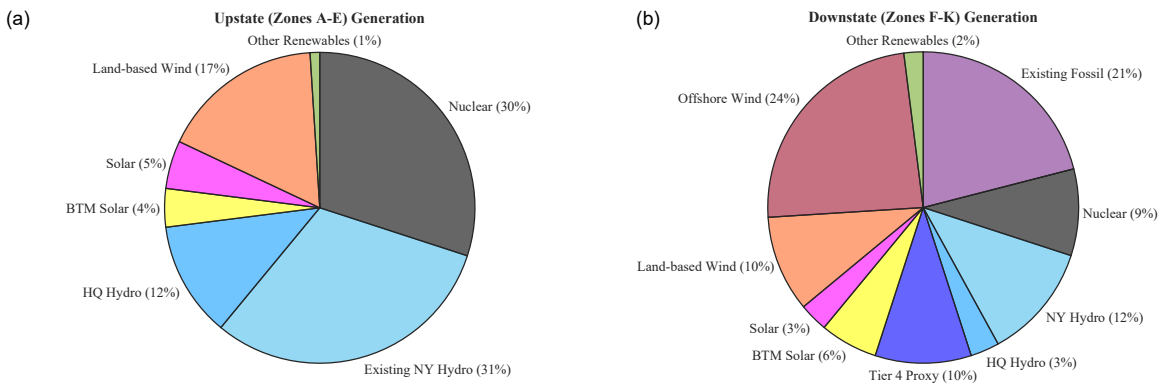

**Figure 2:** Generation shares to meet load 2030 in (a) upstate (zones A to E) and (b) downstate (zones F to K).

The baseline grid for our case study is derived from the NYS power system representation in [44], with modifications to align with the CLCPA 2030 targets. The grid includes 57 buses and 94 transmission lines, including 46 buses in NYS and 9 in neighbouring Independent System Operators (ISOs). The modifications include: (i) six BSS technologies available for optimal allocation of 2030 storage targets, (ii) integration of additional onshore/offshore WTs, UPVs, and DPVs into the baseline grid, as well as upscaling load to achieve the upstate and downstate targets, as shown in Figure 2; (iii) retirement of FFG and NPG based on the 2030 schedule; and (iv) updated transmission line capacity limits for the 11 NYS zones and 3 neighbouring ISOs, reflecting NYS Power Authority transmission projects. These modifications are further explained below.

### 4.1. Energy Storage Technology

We include six scalable BSS technologies—NaSB, LiB, LAB, ZnBrB, VRFB, and Sodium Metal Halide Batteries (ZEBRA) (see Table 4)—as projected for 2030 by the Pacific Northwest National Laboratory [40]. Each BSS is characterized by detailed technical attributes, including *RTE*, *Charging/Discharging Efficiencies*, maximum *DOD*, *Self-Discharge* per day, *Annual RTE Degradation Factor*, cycles at 80% DOD, and *Operational Lifetime*, along with economic characteristics. The combination of these technologies and various storage durations, which influence capital and O&M costs, results in a total of 25 distinct battery types considered in our analysis. A comprehensive list of these 25 BSS types, along with their technical and economic characteristics, is provided in Appendix A.

Table 4: Technical and economic characteristics of BSS technologies applied in this paper

| | | BSS Technology | NaSB | LiB | LAB | ZEBRA | ZnBrB | VRFB |
|---|---|---|---|---|---|---|---|---|
| Economic Char. | Capital Cost | Energy Capacity ($/kWh) | 465 | 189 | 220 | 482 | 192 | 393 |
| | | Power Conversion System ($/kW) | 211 | 211 | 211 | 211 | 211 | 211 |
| | | Balance of Plant ($/kW) | 95 | 95 | 95 | 95 | 95 | 95 |
| | | Construction and Commissioning ($/kWh) | 127 | 96 | 167 | 110 | 164 | 180 |
| | Replacement Cost ($/kWh) | | 199.8 | 409.59 | 190.92 | 202.02 | 216.45 | 144.3 |
| | Fixed O&M Cost ($/kW-yr) | | 3.96 | 7.59 | 3.74 | 6.05 | 4.73 | 9.35 |
| | Variable O&M Cost ($/MWh) | | 1.98 | 2.31 | 0.407 | 0.66 | 0.66 | 0.99 |
| Technical Char. | Storage Duration (Hours) | | 2-8 | 2-8 | 2-8 | 2-8 | 2-10 | 2-12 |
| | RTE | | 0.75 | 0.85 | 0.72 | 0.83 | 0.72 | 0.7 |
| | Discharge Efficiency | | 0.85 | 0.85 | 0.85 | 0.85 | 0.78 | 0.7 |
| | Maximum DOD | | 0.9 | 0.8 | 0.65 | 0.9 | 1 | 1 |
| | Self-Discharge per Day (%) | | 0.05 | 0.2 | 0.08 | 0.3 | 0 | 0.15 |
| | Annual RTE Degradation Factor (%) | | 0.34 | 0.5 | 5.4 | 0.35 | 1.5 | 0.4 |
| | Cycling Limit at 80% DOD | | 4000 | 3500 | 900 | 3500 | 3500 | 10000 |
| | Operational Lifetime (Years) | | 13 | 10 | 3 | 12 | 10 | 15 |

### 4.2. Renewable Energy Sources

To achieve the 2030 NYS renewable generation targets, as outlined in Table 5, we expand the existing wind and solar generation infrastructure. This includes integrating additional onshore and offshore UWTs based on existing and under-construction wind site projects [45], and the National Renewable Energy Laboratory (NREL) WIND Toolkit [46], as well as new DPV and UPV sites informed by the NREL Solar Integration National Dataset [47]. Hourly power generation from these wind and solar sites is modelled using their nameplate capacity and weather variables including wind speed, solar radiation, and temperature, following the method detailed in [32]. In addition to wind and solar, our RES generation includes weather-induced hydropower from two major hydropower plants: the Robert Moses Niagara and the Moses-Saunders hydropower plants. Hydropower data for these plants is sourced from Steinschneider's group [48]. Generation from the remaining small hydropower plants in NYS is aggregated and modelled as a negative load, maintaining current capacities.

Table 5: Zonal RES capacity

| Resource (MW) | Zones | | | | | | | | | | | Total |
|---|---|---|---|---|---|---|---|---|---|---|---|---|
| | A | B | C | D | E | F | G | H | I | J | K | |
| Land-Based Wind | 2692 | 390 | 1923 | 1935 | 1821 | - | - | - | - | - | - | 8761 |
| Offshore Wind | - | - | - | - | - | - | - | - | - | 6391 | 2609 | 9000 |
| UPV solar | 5748 | 656 | 3585 | - | 2268 | 4661 | 2636 | - | - | - | 77 | 19631 |
| DPV solar | 704 | 218 | 596 | 69 | 673 | 827 | 684 | 61 | 90 | 672 | 846 | 5439 |
| Hydro | 2675 | 64 | 109 | 915 | 376 | 270 | 76 | - | - | - | - | 4485 |

### 4.3. Thermal Power Generators

We include three NPGs in the 2030 NYS power system: Nine Mile I (629 MW), Nine Mile II (1299 MW), and Fitzpatrick (854.5 MW). Although Ginna (581.7 MW) is currently operational, it is scheduled for permanent shutdown in 2029 and is therefore excluded from our model. It is assumed that each NPG operates at its maximum allowed generation limit throughout the entire operation period.

The baseline grid for our case study includes 227 FFGs in the NYS regions, categorized into five groups: (I) 100 combustion turbines, ranging from 32 MW to 102 MW; (II) 48 combined-cycle gas turbines, with capacities between 270 MW and 1220 MW; (III) 35 jet engines, producing 35 MW to 60 MW; (IV) 31 steam turbines, with outputs from 110 MW to 1000 MW, and (V) 13 internal combustion engines, generating up to 10 MW. To meet the 2030 upstate and downstate generation targets (see Figure 2), we retire all FFGs in the upstate region (zones A to E). In the downstate (zones F to K), 21% of the annual load is expected to be supplied by FFGs. To achieve this, we developed a separate optimization model (detailed in Appendix B) that selects a subset of currently operational FFGs, minimizing total generation costs, $CO_2$ emissions, and load shedding, while maximizing utilization rates and adhering to transmission constraints. The optimization results indicate a substantial decrease in FFG capacity, achieving an overall reduction of 22 GW. Following the retirement process, 42 FFGs remain operational—14 within NYS and 28 in neighbouring ISOs.

### 4.4. Transmission Lines
The baseline NYS power system grid includes 94 transmission lines, with 68 connecting the 11 zones within NYS and 8 linking NYISO to neighbouring ISOs, including IESO, ISONE, and PJM. Transmission limits are updated based on the NYISO's Reliability Needs Assessment report [49]. Additionally, We incorporate a new HVDC line from Zone D to Zone J and two new transmission lines to integrate offshore wind energy into the statewide grid, as outlined in future NYS transmission projects planned for 2030 [50].

### 4.5. Electricity Load
We estimate hourly zonal loads in NYS using a random forest model trained on 2005–2019 data [51], with predictors including day of the week, day of the year, and the last 24 hours of average temperature. Predicted loads are bias-corrected using quantile mapping to preserve variance. Historical predictions (1980–2004) are merged with actual data (2005–2019) to create a continuous hourly load record, which is scaled to meet the 2030 summer and winter peak load forecasts [52].

### 4.6. Modelling Spatio-temporal Co-variability in Load and RES Generation Scenarios
Climate variables, including solar radiation, temperature, wind speed, and river dynamics, were collected across NYS for 1980 to 2019 [51]. These variables are used in separate models to generate wind, solar, and hydropower outputs, as well as electricity load. The 40 years of simulated data are treated as scenarios in the stochastic optimization model, with load and RES generation estimated at an hourly resolution. This approach captures the spatio-temporal co-variability between resources while simulating long-term patterns over four decades. Details of the uncertainty of renewables and loads model is presented in Appendix C.

## 5. Computational Results and Insights
The SAA is implemented to optimize the size, type, and location of BSS technologies for the 2030 NYS power system case study, with the objective of minimizing total operational and planning costs. While the data used are specific to the NYS power system, the analysis and insights can be readily applied to other power systems.

### 5.1. Analysing Optimal BSS Allocation in 2030 NYS Renewable Target

In this section, we present the results of the optimal allocation of 3000 MW *Rated Power Capacity* of BSS for the year 2030, which serves as the *Main Strategy* of our paper.

In the SAA algorithm, the proposed model is solved using Gurobi on a computer equipped with an Intel Core i7-13700H (2.40 GHz) processor and 16 GB of RAM. We consider $\alpha = 0.05$, $M = 40$, and $W = 20$, and implement the SAA algorithm for various values of $M'$. The results of the SAA, including $\mu_U$, $\sigma_U$, the 95% CI for the LB, and the 95% CI for the UB, are reported in Table 6.

Also in Table 6, the Gap (LB–UB)% indicates the difference between the lower bound of the LB confidence interval and the upper bound of the UB confidence interval, representing the range within which the true objective value lies with 95% confidence. The table highlights that the Gap (LB–UB)% decreases as the number of scenarios $M'$ increases. However, the problem becomes unsolvable when the number of scenarios exceeds 7. Since a gap of 11.4% is reasonable, we adopt $M' = 7$ for the remainder of the paper.

In Table 7, we report the number of variables and constraints in our optimization problem, as well as the average CPU time and optimality gaps for solving the problem under different values of $M'$ in the SAA implementation.

Table 6: Statistical Results for Different Values of $M'$

| $M'$ | $\mu_U$ | $\sigma_U$ | CI for LB | CI for UB | Gap (LB–UB) (%) |
|---|---|---|---|---|---|
| 3 | $1.6191 \times 10^{10}$ | $0.2402 \times 10^{10}$ | $(1.2935, 1.5578) \times 10^{10}$ | $(1.5446, 1.6953) \times 10^{10}$ | 23.7% |
| 5 | $1.6169 \times 10^{10}$ | $0.2567 \times 10^{10}$ | $(1.3814, 1.5831) \times 10^{10}$ | $(1.5373, 1.6964) \times 10^{10}$ | 18.5% |
| 7 | $1.6166 \times 10^{10}$ | $0.2436 \times 10^{10}$ | $(1.4986, 1.6889) \times 10^{10}$ | $(1.5411, 1.6920) \times 10^{10}$ | 11.4% |

Table 7: Model characteristics across $M'$ selected scenarios

| Model Characteristics | $M' = 1$ | $M' = 3$ | $M' = 5$ | $M' = 7$ |
|---|---|---|---|---|
| Number of Constraints | 5,203,571 | 15,610,567 | 26,017,650 | 36,424,792 |
| Number of Continues Variables | 1,743,338 | 5,229,834 | 8,716,451 | 12,203,127 |
| Average CPU Time (Seconds) | 544.343 | 1,481.797 | 3,272.922 | 5,197.531 |
| Average Optimality Gap | 4.05312e-06 | 6.1035e-05 | 2.3365e-04 | 4.05176e-04 |

To evaluate the impact of stochasticity on the optimal allocation of BSS, two well-known metrics, the Expected Value of Perfect Information (EVPI) and the Value of the Stochastic Solution (VSS), are calculated for the *Main Strategy*. The VSS quantifies the difference between the objective value of the two-stage stochastic program, referred to as the recourse problem (RP), and the expected value of deterministic objectives across different scenarios, denoted as EEV. In this case, the first-stage decisions are determined by solving the problem with the expected value of the stochastic parameters and are fixed in the problem.

EVPI, on the other hand, represents the maximum potential benefit a decision-maker could achieve if precise and complete information about future stochastic parameters were available. In stochastic optimization, EVPI is computed as the difference between the RP value and the wait-and-see (WS) solution, which is the expected value of solutions obtained by solving the problem separately for each scenario to determine both first- and second-stage decisions. Further details on the definitions and calculations of EVPI and VSS are provided in Appendix D.

Table 8 presents the RP, WS, EVPI, EEV, and VSS values for the *Main Strategy* with 3000 MW *Rated Power Capacity*. As shown in Table 8, the significant VSS value, approximately 22% of the RP value, underscores the importance of accounting for the uncertainties in load and maximum renewable generation when determining the optimal allocation of BSS.

Table 8: Values of RP, WS, EVPI, EEV, and VSS for the *Main Strategy* with 3000 MW *Rated Power capacity*

| RP | WS | EVPI | EEV | VSS |
|---|---|---|---|---|
| $1.6166 \times 10^{10}$ | $1.6147 \times 10^{10}$ | $1.8915 \times 10^{7}$ | $1.9728 \times 10^{10}$ | $3.5613 \times 10^{9}$ |

Figure 3 shows the decisions regarding the location, capacity, type, and duration of the allocated BSS technologies in each zone for the *Main Strategy* with 3000 MW *Rated Power capacity*. As shown in Figure 3, zones C, D, and E account for 60.5% of the BSS capacities, while zones J and K account for the remaining 39.5%. Besides, zone C has the highest share at 34.5%, and zone D has the lowest share at less than 2%.

Despite its low *RTE*, NaSB technology, with a maximum *Storage Duration* of 8 hours, is selected in zones C, D, and E due to its favourable technical and economic characteristics compared to other technologies. This indicates that zones C, D, and E prioritize deploying an economical technology with appropriate technical features. Also, ZEBRA technology, with its maximum *Storage Duration* of 8 hours, is chosen in zones J and K because of its high *RTE* and *Discharge Efficiency*. Also, ZnBrB technology, which offers a maximum *Storage Duration* of 10 hours, is allocated in zone J due to its low capital cost and extended *Storage Duration*.

Given that the variability of offshore wind power exceeds that of onshore wind and solar power, ZEBRA technology is selected for zones J and K, which have a higher share of offshore wind. This choice enables more frequent charging and discharging to adapt the variability of offshore wind compared to other zones.

Figure 3 also shows that no BSS is allocated to zones A, B, and F through I. These results are further interpreted by analysing the average operational decisions of the stochastic program, as follows:

- Zone A, which has a high level of hydropower generation from the Robert Moses Niagara unit, experiences a surplus of generation relative to its load. This surplus energy is continuously transferred to Zone B via transmission line A-B, which operates without congestion. As a result, no BSS capacity is allocated to Zone A. Similarly, Zone B does not have any allocated capacity because it can reliably receive power from Zone A. Additionally, an on-grid NPG in Zone B consistently generates power to meet the base load demand within the zone.
- Zone F has the lowest average curtailed RES in the downstate area, indicating that most of its renewable generation is either utilized locally or transferred to other zones. Power transmission is feasible as there is no congestion in the transmission lines F-G, G-H, and H-I, according to our results.
- Zones G, H, and I have less than 10% RES penetration, making it suboptimal to install any BSS in these zones. Besides, zones H and I have high demand compared to their RES generation. To address this, the NPG in zone H consistently generates power to supply the base load in zones G-I.

*Managerial Insight I*: BSS technologies are most valuable in zones with high RES generation relative to demand, particularly when transmission capacity to neighbouring areas is constrained. Our results indicate that such zones are prime candidates for BSS deployment, as stored energy can help balance local surplus generation. Nevertheless, when multiple interconnected zones experience no transmission congestion yet share a high-RES-to-load profile, each can still benefit from BSS installation, as surplus energy may be stored locally and subsequently utilised to support neighbouring zones during peak demand periods.

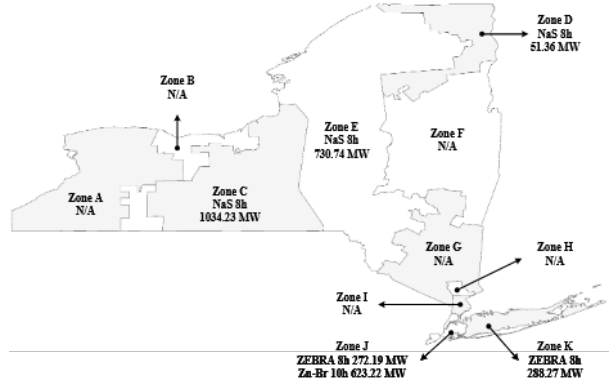

**Figure 3:** Allocated BSS in each zone for the *Main Strategy* with 3000 MW rated power capacity.

**Cycling Number Analysis**. The annual cycling number of the allocated BSS in the *Main Strategy* demonstrates how diverse BSS technologies operate within the power system. The annual cycling number of allocated NaSB technology in zones C, D, and E reaches 96% of their maximum allowed value. In contrast, the ZEBRA technology achieves 100% and 98% of their maximum cycling limit in zones J and K, respectively. Additionally, ZnBrB technology with almost 150 annual cycles is operated in zone J because the cycling limit of the ZEBRA in this zone has been fully utilized.

*Managerial Insight II*: BSS technology choice, size, and location depend on the zonal generation–load profile and transmission capacity. In BSS planning, these system characteristics strongly influence the preferred technology type, while also affecting optimal capacity and placement. Additionally, the energy source mix shapes technology selection; for example, zones with high offshore wind penetration may benefit from BSS options with higher RTE and discharge efficiency to better match the variability and output profile of wind generation.

**Battery Costs Analysis**. The annual capital, replacement, and fixed O&M costs of BSS, as determined in the first stage of the stochastic problem for the *Main Strategy*, are 1154.95, 328.06, and 24 billion USD, respectively. The costs associated with variable O&M and degradation of BSS are less than 2% of the total annual costs in the NYS. Additionally, the annual $CO_2$ emissions cost exhibits the highest variability across scenarios, as it depends on power generation. This variability primarily originates from steam turbine generators. More details about the annual costs for operating FFG, $CO_2$ emissions, variable O&M of BSS, and BSS degradation are presented in Appendix E.

**Power Balance Analysis**. Figure 4 highlights the power balances in the optimized system operation under the *Main Strategy*, illustrating summer and winter weeks with the maximum load shedding. The variables are defined as follows: $P_{(t)}^{NPG}$ and $P_{(t)}^{FFG}$ represent the average power generation of NPG and FFG in NYS, respectively; $P_{(t)}^{HP}$ is the total average power generation from BHD and SHD; $P_{(t)}^{WT}$ is the average power generation from UWT; $P_{(t)}^{PV}$ is the total average power generation from UPV and DPV in NYS; Positive values of $P_{(t)}^{BSS}$ represent the average discharged power of BSS, while negative values represent the average charged power of BSS in NYS; Positive values of $P_{(t)}^{FL}$ indicate the average power imported from neighbouring ISOs to NYS, while negative values indicate power exported from NYS to neighbouring ISOs; $P_{(t)}^{LS}$ represents the average load shedding in NYS; and $P_{(t)}^{PD}$ represents the average load demand in NYS.

As shown, the average PV generation is higher in summer, whereas the average WT generation is higher in winter. The FFG output during peak load hours is greater in summer because the NYS power system receives additional power support from neighbouring grids. In winter, the NYS power system attempts to export surplus wind generation to ISONE during off-peak hours. Furthermore, the charging and discharging of BSS are more pronounced in summer, as the load shedding is higher during peak load hours. These figures show that BSS technologies charge during off-peak hours when RES generation is high and discharge during peak load hours. This daily pattern of BSS operation is consistently observed in both summer and winter.

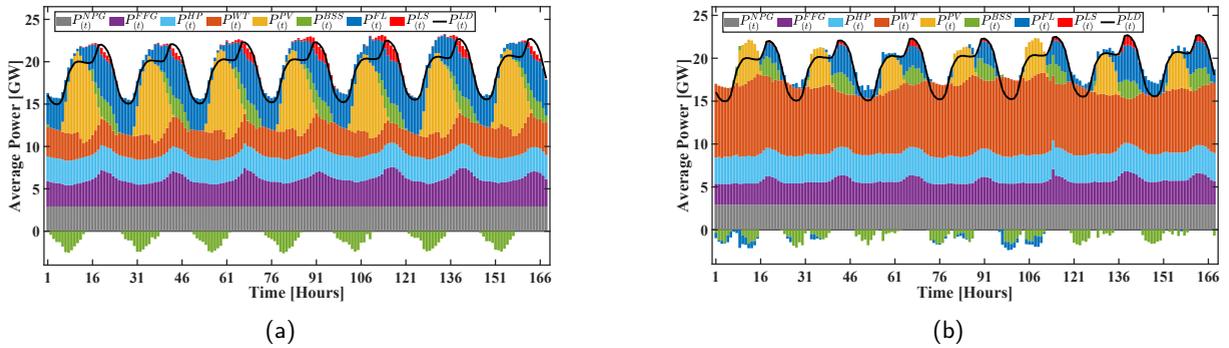

**Figure 4:** Power balance in a week with maximum load shedding in *Main Strategy* for (a) summer and (b) winter.

To analyse the impact of BSS on power system operations in terms of power generation and load, we calculated the normalized load. This involves dividing the load at each time step by the maximum load observed over the entire period, resulting in a scaled metric ranging from 0 to 1. Figure 5 presents the normalized load profile for the week with the highest load shedding observed during the year, comparing the load of the *Main Strategy* with 3000 MW *Rated Power Capacity* of BSS to the load with no BSS.

The load profile is influenced by BSS in two distinct ways: during charging, BSS acts as a positive load by drawing additional power from the grid, while during discharging, BSS acts as a negative load by supplying power back to the grid. As shown in Figure 5, the load profile without BSS is more volatile, exhibiting sharper increases and decreases between peak and off-peak periods. This represents the natural variation in grid demand without any load-smoothing mechanism. With BSS, the peaks are smoothed, and the troughs are filled as the system stores excess energy during periods of low demand and discharges it during high demand. This results in a load profile that is less volatile and more stable compared to the case without BSS.

### 5.2. Comparison of Our Results with NYISO Guidelines for BSS Allocation

To emphasize the improvements achieved through the optimal BSS allocation, the results of the *Main Strategy* are compared to those of the optimal solution based on NYISO guidelines provided in [52]. In the optimization model following the NYISO guidelines, pre-defined capacities are allocated to each zone as specified in Table 9. Table 9 also indicates the decisions regarding the type and duration of the allocated BSS technologies in each zone based on the NYISO guidelines.

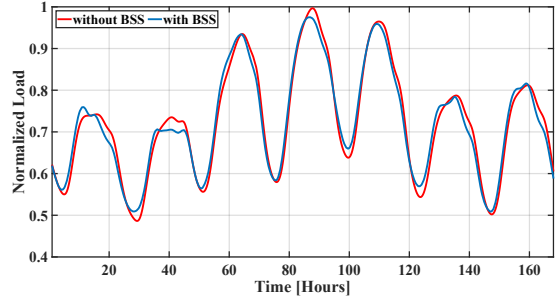

**Figure 5:** Normalized load profile of the *Main Strategy* with 3000 MW rated power capacity of BSS and no BSS.

Table 9: Assumed zonal capacity and allocated type of BSS based on the NYISO guidelines

| Zone | A | B | C | D | E | F | G | H | I | J | K |
|---|---|---|---|---|---|---|---|---|---|---|---|
| Capacity | 150 MW | 90 MW | 120 MW | 180 MW | 120 MW | 240 MW | 100 MW | 100 MW | 100 MW | 1320 MW | 480 MW |
| BSS Type | NaSB 6h | NaSB 8h | NaSB 8h | NaSB 8h | NaSB 8h | ZEBRA 8h | ZEBRA 8h | ZEBRA 8h | ZEBRA 8h | ZEBRA 6h | ZEBRA 8h |

As shown in Table 9, NaSB technology is optimally selected for the upstate area, while ZEBRA technology is selected for the downstate area. The NYISO guidelines result in a more extensive distribution of BSS capacities across zones compared to the optimal solution in the *Main Strategy*. Consequently, the BSS requires longer discharge time, leading to the selection of BSS technologies with high discharge durations at rated power, high RTE, and high discharge efficiency under this strategy.

Table 10 compares the optimal values of the first-stage and average second-stage costs, as well as the average load shedding, between the *Main Strategy* and the optimal solution based on the NYISO guidelines across 40 scenarios. The results clearly show that the optimal BSS allocation in the *Main Strategy* reduces the total cost by 0.3% and load shedding by 3.4%, respectively.

Table 10: Comparison of the *Main Strategy* and the NYISO guidelines

| Strategy | BSS Costs (MM$) | | | | | FFG Costs (MM$) | | Total Cost (MM$) | Load Shedding (GW) |
|---|---|---|---|---|---|---|---|---|---|
| | Capital | Replacement | Fixed O&M | Variable O&M | Degradation | Operating | $CO_2$ Emission | | |
| *Main Strategy* | 1154.94 | 328.06 | 24 | 3.64 | 20.63 | 680.29 | 2274.46 | 4486.03 | 567.81 |
| NISO Guidelines | 1192.76 | 291.97 | 24 | 3.83 | 13.66 | 684.46 | 2289.08 | 4499.76 | 587.79 |
| Difference | -3.17% | 12.36% | 0% | -4.96% | 51.02% | -0.6% | -0.64% | -0.3% | -3.4% |

## 5.3. The impact of BSS on the Power System with High RES Penetration

In this subsection, we analyse the impact of BSS allocation on the power system, focusing on load shedding and transmission congestion. Figure 6 illustrates the reduction in transmission congestion rates across various transmission lines between zones and the decrease in load shedding during the summer and winter seasons of the NYS power system. This comparison is based on the *Main Strategy* with 3000 MW *Rated Power Capacity* of BSS versus the strategy without BSS. We define the transmission congestion rate as the fraction of time during which the power flow on a transmission line reaches its maximum limit. Load shedding is evaluated using three metrics: Figures 6(a) and 6(b) display the total yearly load shedding reduction for summer and winter, respectively; Figures 6(c) and 6(d) depict the maximum load shedding reduction observed in summer and winter, respectively; Figures 6(e) and 6(f) present the reduction in the total number of hours with load shedding during summer and winter, respectively. Across all scenarios, the size of the marker (dot) indicates the average intensity of the load shedding metrics in each zone, while the colour of the lines represents the average likelihood of transmission congestion between zones. A green line signifies a reduction in transmission congestion rates, while a red line indicates an increase.

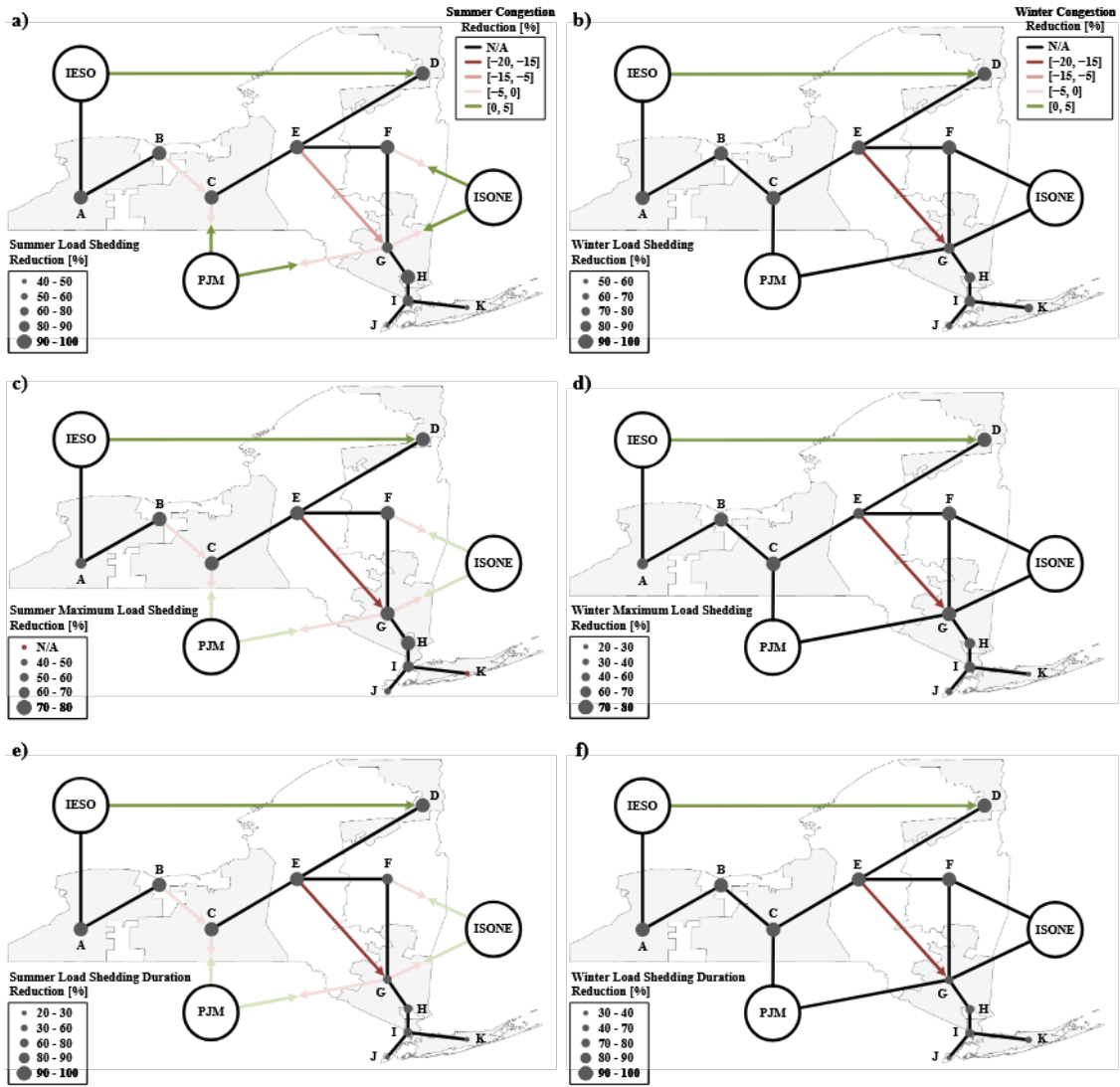

**Figure 6:** Reduction in transmission congestion rate and three different load shedding metrics in summer and winter, (a) summer load shedding reduction, (b) winter load shedding reduction, (c) summer maximum load shedding reduction, (d) winter maximum load shedding reduction, (e) summer load shedding duration reduction, and (f) winter load shedding duration reduction.

Notably, the results for the power system without BSS align with findings by [44] and [32], where the main transmission congestions occur in: (i) the transmission lines B→C and E→G in the upstate area, and (ii) the transmission line between NYS and neighbouring ISOs. Our results underscore the significant impact of BSS on reducing load shedding, with greater reductions observed in the upstate area (zones A to E) compared to the downstate area (zones F to K) during both summer and winter. However, the effect of BSS on transmission congestion varies across lines and depends on the specific characteristics of the power system. Changes in transmission congestion on these lines are generally minimal, except for the transmission line E→G, where transmission congestion increases following BSS allocation. This transmission line, which links the upstate and downstate areas, experiences congestion because the additional BSS capacity installed in the upstate area stores surplus renewable energy near generation sites, thereby increasing power flows on already constrained lines and further exacerbating congestion.

Our findings show that although BSS can reduce curtailment locally, they do not necessarily alleviate congestion across the broader transmission system when located close to generation sites, and in some cases may even increase it. Overall, the results suggest that transmission line capacity is a key factor in deciding where to site BSS, but the impact on congestion after deployment depends strongly on the specific characteristics of the system.

**RES Utilization Analysis**. The utilization of hydro, solar, and wind power increases by an average of 2.5%, 3.8%, and 2%, respectively, when BSS is added to the system, leading to a reduction in RES curtailment. However, the NYS power system is unable to fully accommodate the total renewable generation due to transmission congestion in the lines B-C and E-G. As a result, approximately 3.6% of the RES power generation is curtailed. More details about the RES utilization factors are presented in Appendix F.

*Managerial Insight III*: BSS integration can reduce load shedding and RES curtailment, but benefits are limited in high-load regions with constrained transmission capacity. Although integrating BSS into power systems significantly reduces these issues, regions with high load relative to generation may not fully benefit when interconnecting transmission lines are limited. BSS can mitigate curtailment locally by storing surplus renewable energy near generation sites; however, system-wide transmission congestion still restricts further reductions as BSS capacity grows. To fully realize BSS benefits, congestion must be addressed, either by upgrading or adding transmission lines or by adopting power-to-X solutions (e.g., hydrogen) that bypass the transmission network.

## 5.4. Impact of BSS Capacity Increase on Energy Transition

In this section, we aim to address two key questions: (I) How much BSS capacity is required to eliminate load shedding and RES curtailment in the system? (II) Can we fully utilize RES by increasing BSS capacity? To provide a detailed analysis, we present the simulation results for two additional strategies with BSS *Rated Power Capacities* of 4000 MW and 5000 MW. Figure 7 illustrates the decisions regarding the location, capacity, type, and duration of the allocated BSS technologies in each zone for strategies with *Rated Power Capacities* of (a) 4000 MW and (b) 5000 MW, respectively.

Similar to the *Main Strategy* with a 3000 MW *Rated Power Capacity*, no BSS is allocated in zones A and F through I. This indicates that increasing BSS capacity does not affect the optimal locations of the BSS technologies. Additionally, only NaSB is selected in the upstate area, while ZEBRA and ZnBrB are exclusively selected in the downstate area. Even with a 66% increase in allocated *Rated Power Capacity* of BSS in the 5000 MW strategy, the choice of BSS technologies remains unchanged.

Regarding the selected types of BSS for the NYS power system in these strategies, while LiBs are the most commonly used BSS technology in studies and applications, they are not selected in the optimal BSS allocation. Although LiB technology offers high RTE and a low self-discharge rate, it also has a high replacement cost and a low maximum DOD. As a result, its superior technical performance cannot fully compensate for its economic disadvantages. Similarly, LABs are not chosen due to their lowest RTE, maximum DOD, and lifespan among the available BSS technologies. VRFBs, on the other hand, offer the highest discharge duration—up to 12 hours—enabling the system to operate at rated power for extended periods compared to other technologies. Despite these advantages, including a high cycling limit and maximum DOD, VRFB technology is not selected because of its low RTE and discharge efficiency.

In conclusion, the system prioritizes BSS technologies that balance technical and economic characteristics. While technical features significantly influence optimal charging and discharging operations, economic factors play a critical role in determining the capital costs of BSS installation.

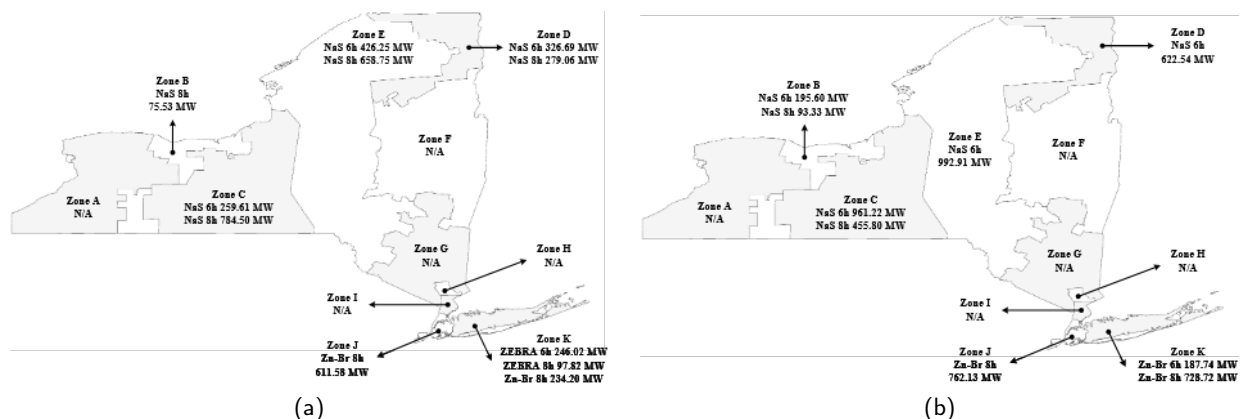

**Figure 7:** Allocated BSS in each zone for (a) 4000 MW and (b) 5000 MW *Rated Power Capacity*.

### 5.4.1. Impact of BSS Capacity on Load Shedding and RES Curtailment

This section analyses the impact of BSS capacity on load shedding and RES curtailment. First, the optimization model is solved after relaxing the BSS capacity constraint (Constraint 34). The results indicate that a rated BSS capacity of 6740 MW is required to minimize load shedding in the NYS power system.

Figure 8 illustrates the relationship between BSS capacity and (a) annual load shedding and (b) annual RES curtailment. As shown in Figure 8(a), increasing BSS capacity reduces load shedding. However, load shedding saturates at 209 GWh due to transmission congestion, rendering additional capacity beyond 6740 MW ineffective. For instance, solving the model with a maximum *Rated Power Capacity* of 7000 MW yields the same load shedding outcome as a capacity of 6740 MW. Figure 8(b) demonstrates that increasing BSS capacity reduces RES curtailment. BSS absorbs surplus energy, allowing for greater utilization of renewable energy sources. As BSS capacity increases, it enables the storage of more excess renewable energy, reducing both curtailment and load shedding. By storing energy during periods of surplus RES generation and discharging it during times of high demand or low generation (e.g., cloudy or windless periods), BSS enhances grid flexibility, buffering fluctuations in RES generation against stable electricity demand and improving the effective utilization of RES.

Furthermore, when RES is generated far from demand centres, the transmission infrastructure may lack the capacity to transport the full output. In such cases, BSS located near RES generation sites can store excess energy until the transmission system becomes less congested, mitigating both load shedding and curtailment.

As shown in Figure 8, increasing BSS capacity reduces both load shedding and RES curtailment. However, the efficiency of BSS in mitigating these issues diminishes as BSS capacity increases. Accordingly, LSRCE and RCRCE metrics are illustrated in Figures 9(a) and 9(b), respectively. Based on the definitions of LSRCE and RCRCE, as provided in Equations (51) and (52), higher BSS allocation costs combined with smaller reductions in load shedding and RES curtailment lead to lower LSRCE and RCRCE values. A higher LSRCE indicates greater cost-effectiveness in reducing load shedding, while a higher RCRCE signifies more cost-effective reductions in RES curtailment.

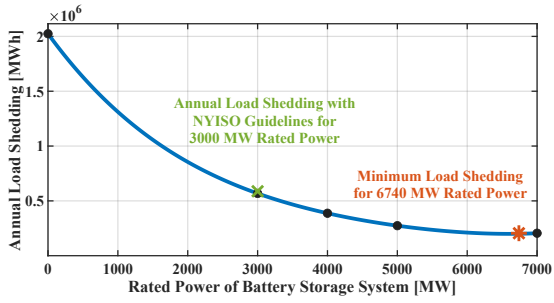
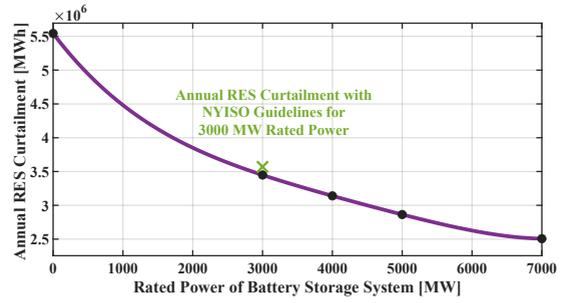

**Figure 8:** (a) annual load shedding and (b) RES curtailment for different rated power capacities of BSS.

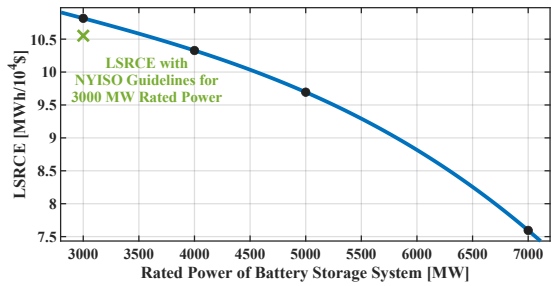
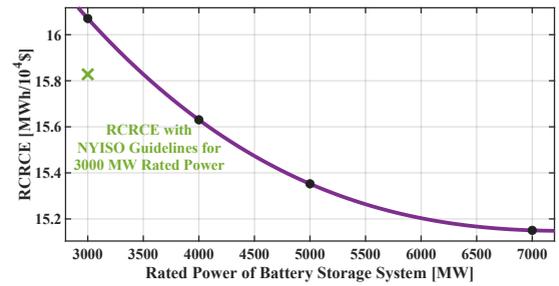

**Figure 9:** (a) LSRCE and (b) RCRCE for different *Rated Power Capacities* of BSS.

Figure 9 illustrates that both LSRCE and RCRCE decrease non-linearly as BSS capacity increases. Although load shedding and RES curtailment are reduced, the cost of achieving these reductions rises significantly, resulting in a sharp decline in LSRCE and RCRCE. This trend is influenced by the non-linear characteristics of the power system. Initial additions of BSS capacity are highly effective at managing peak demand, providing substantial reductions in load shedding and RES curtailment. However, as BSS capacity increases, the effectiveness of additional capacity diminishes due to grid constraints, transmission congestion, and the inherent variability of RES. These factors limit further reductions in load shedding and RES curtailment, causing the cost-effectiveness metrics to decline.

*Managerial Insight IV*: BSS is vital for mitigating RES curtailment and load shedding, but its cost-effectiveness declines as capacity grows. In a renewable-dominant power system, BSS plays a crucial role in reducing these issues, yet it cannot fully utilize all curtailed RES or entirely eliminate load shedding. As BSS capacity increases, cost efficiency drops sharply, reflecting diminishing returns on investment. Early investments deliver substantial benefits, but beyond a certain capacity threshold, the marginal gains from each additional unit diminish progressively. Policymakers must therefore determine the optimal BSS capacity, weighing it against other strategies such as transmission upgrades or demand-side management to enhance overall system efficiency.

### 5.4.2. Impact of BSS Capacity on Renewable Energy Penetration

In this subsection, the impact of BSS capacity on RES penetration is evaluated. RES penetration is defined as the ratio of energy generated from RES to the total energy consumed within the system. Figure 10 illustrates that average RES penetration increases gradually with higher rated BSS capacities, eventually saturating at 66% of total load demand. It is obvious that larger BSS capacities reduce RES curtailment, enabling a greater proportion of renewable energy to be utilized. However, once system flexibility is maximized through BSS allocation, surplus renewable energy during off-peak hours remains underutilized due to transmission congestion, limiting further improvements in RES penetration. As a result, achieving the NYS power system's 70% RES target by 2030 is unattainable with BSS allocation alone. Addressing transmission congestion or expanding RES generation capacity is essential to meet this target.

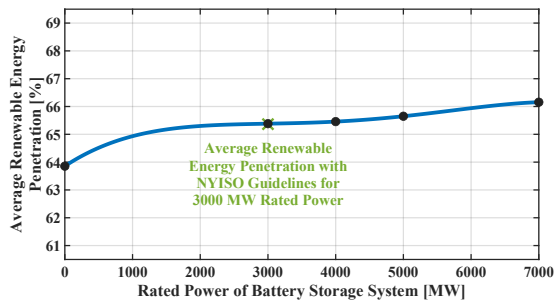

**Figure 10:** Average renewable energy penetration for different rated power capacities of BSS.

### 5.5. Analysing the Impact of RES Penetration on BSS Efficiency

Based on existing literature, achieving low or zero emissions targets in a power system while avoiding load shedding involves five main strategies: (I) BSS allocation, (II) transmission line expansion, (III) utilizing alternative energy sources, such as hydrogen, to make use of curtailed RES, (IV) expanding RES generation capacity, and (V) load response management.

In this subsection, we analyse how changes in RES generation capacity affect the power system and the effectiveness of the BSS allocation strategy. In this study, RES generation data spanning 40 years is used to create 40 scenarios for the optimization problem. These scenarios reflect varying levels of RES generation and penetration within the system. For each scenario, the power system is simulated both without BSS and with the optimal allocation of 3000 MW *Rated Power Capacity* of BSS.

Figure 11 shows (a) annual load shedding and (b) annual RES curtailment for each RES penetration scenario. The results indicate that annual load shedding significantly decreases as RES penetration increases, while annual RES curtailment rises with higher RES penetration. This highlights a different mechanism for managing a renewable-dominant system compared to the BSS allocation strategy. For example, as RES penetration increases from approximately 62% to 69%, annual load shedding decreases from around 900 GWh to 200 GWh. Conversely, annual RES curtailment increases from roughly 2200 GWh to 4500 GWh.

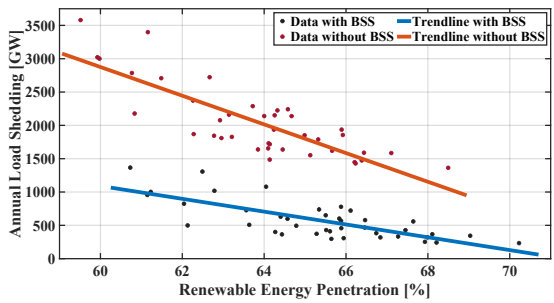 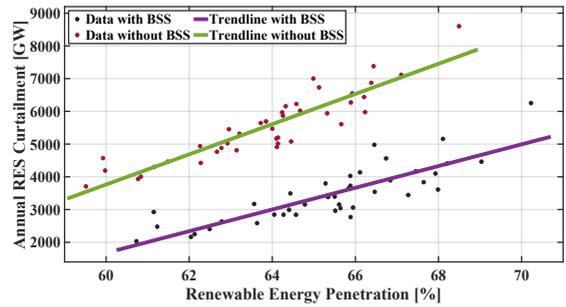

**Figure 11:** (a) annual load shedding and (b) annual RES curtailment per RES penetration for each scenario.

The relationship between RES penetration and curtailment highlights a broader challenge in power systems with high shares of RES. When RES generation exceeds load and transmission capacity is congested, surplus energy cannot be fully utilized by the system, leading to curtailment. In such cases, strategies like BSS allocation can play a significant role by absorbing and utilizing this curtailed energy.

Figure 12 illustrates the reduction in load shedding after the allocation of 3000 MW *Rated Power Capacity* of BSS across different RES penetration scenarios. The results emphasize the increasing effectiveness of BSS in reducing load shedding as RES penetration grows.

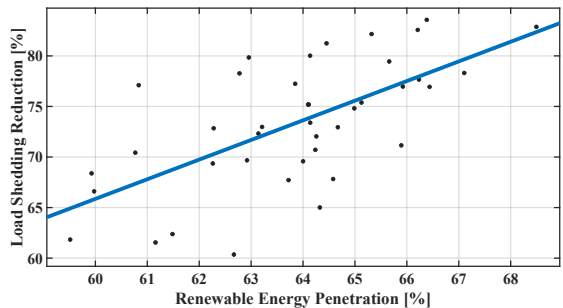

**Figure 12:** Load shedding reduction per RES penetration for each scenario.

*Managerial Insight V*: BSS allocation supports higher RES penetration by reducing curtailment, which in turn lowers load shedding. Increasing RES generation not only increases penetration but also expands the potential role of BSS, as surplus generation that cannot be consumed or transmitted is curtailed. Figure 13 illustrates these relationships, highlighting the interconnected effects of RES generation, curtailment, and BSS allocation. These findings emphasize the importance of a balanced policy approach that integrates RES expansion, transmission upgrades, and BSS deployment to achieve net-zero targets efficiently.

## 6. Conclusion

We propose a novel optimization-based framework for BSS planning in renewable-dominant power systems. Aligned with future energy system targets, our framework models the retirement of fuel-based generators and optimizes the location, capacity, and types of batteries to minimize load shedding and overall system costs. These costs include both the capital investment for battery deployment and operational expenses. To account for renewable generation and load uncertainties, the framework employs a two-stage stochastic programming approach. Computational tractability is ensured through the use of the SAA algorithm.

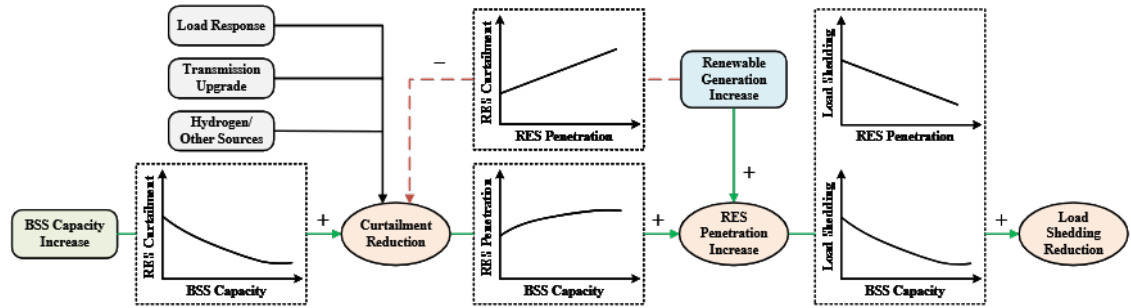

**Figure 13:** Relation chart connecting various factors influencing curtailment, RES penetration, and load shedding.

We apply our framework to the NYS power system as a real-world case study, utilizing 40 years of historical weather data to model uncertainties. Extensive experiments are conducted to analyse the impact of BSS on renewable-dominant power systems. Furthermore, we introduce two metrics to assess the cost-effectiveness of BSS in power systems, providing valuable insights for policymakers. These metrics help determine the optimal BSS capacity while comparing it to alternative strategies for reducing load shedding and RES curtailment.

In conclusion, the optimal BSS allocation strategy with a rated capacity of 3000 MW reduces RES curtailment to approximately 3.6% of total renewable generation, thereby increasing the share of renewable penetration and resulting in a 71.95% reduction in load shedding. Achieving the minimum load shedding would require a BSS capacity of 6740 MW; however, beyond this threshold, additional capacity yields limited benefits due to transmission congestion. At this level, RES penetration saturates at approximately 66% of total demand, underscoring the need for coordinated BSS deployment and transmission network upgrades to maximize renewable integration and overall grid resilience.

## 7. Model Boundaries and Future Work

This study develops a modelling framework to determine the optimal location, size, and technology of BSS while addressing grid constraints and weather variability in a renewable-dominant grid. The framework provides valuable techno-economic insights into spatially resolved BSS deployment, but is bounded by a defined set of assumptions that shape the scope of the results. Recognizing these boundaries is not only important for interpreting the findings accurately but also highlights promising directions for future research.

In our work, the power system case study is represented in an abstracted form. Although it has been validated, factors such as market mechanisms, transmission losses, dynamic ratings, and solar efficiency variability are not yet modelled. Forecasts for renewable generation and demand were carefully validated, but more advanced methods could further improve their accuracy; however, this was not the focus of our work. We also assume that CLCPA targets are achieved by the year 2030, though future work should examine deviations from these targets and their effects on BSS deployment decisions. Such deviations could arise from changes in load patterns, renewable mixes, transmission constraints, or climate change impacts. In addition, neighbouring ISOs were modelled based on 2019 operations, while ongoing grid transitions may alter transfer dynamics. Finally, BSS costs are treated as fixed, without accounting for potential cost reductions from technological learning or supportive policies.

This research opens several avenues for future work. One of them is to apply the proposed framework to other case studies. Another direction is to integrate techno-economic BSS planning with additional strategies for enhancing the power system. For example, transmission capacity expansion or Power-to-X solutions (e.g., converting curtailed renewable energy to hydrogen) could help reduce system congestion. These solutions, when combined with BSS, have the potential to significantly improve overall efficiency and sustainability. Another important avenue is incorporating AC-based post-processing or hybrid DC/AC modelling approaches for short-term operation. This would provide a more physically accurate representation of dispatch results by capturing voltage and reactive power constraints, particularly in load-dense zones. Future work could also include environmental cost analysis of BSS technologies and the integration of life cycle assessment indicators for a more comprehensive evaluation of storage options. Finally, a promising line of research is to examine the impact of climate change on BSS techno-economic planning. This includes developing models that explicitly account for weather variability under different scenarios and applying them to later years, such as 2040 and beyond.

**Note**. Appendices are provided as **supplementary material** and have been submitted as an E-component.